\documentclass{aa}

\usepackage{natbib}
\usepackage{graphics}
\bibpunct{(}{)}{;}{a}{}{,}

\newcommand{\mic}{\hbox{${\mu}$m}}
\newcommand{\Msun}{$M_{\sun}$}
\newcommand{\Rsun}{$R_{\sun}$}
\newcommand{\Lsun}{$L_{\sun}$}
\newcommand{\Mstaro}{$M^0_{\star}$}
\newcommand{\Mstar}{$M_{\star}$}
\newcommand{\Rstar}{$R_{\star}$}
\newcommand{\Lstar}{$L_{\star}$}
\newcommand{\Tstar}{$T_{\star}$}
\newcommand{\dustgas}{$\rho_{\rm d}/\rho$}

\newcommand{\SiCC}{[SiC,\,C]}
\newcommand{\SiCCMgFeS}{[SiC,\,C]+[Mg,\,Fe]S}
\newcommand{\HmK}{\mbox{$H{-}K$}}
\newcommand{\JmH}{\mbox{$J{-}H$}}
\newcommand{\JmK}{\mbox{$J{-}K$}}
\newcommand{\irc}{\object{IRC $\!$+10\,216}}

\defcitealias{Osterbart_etal2000}{Paper~I}
\defcitealias{Men'shchikov_etal2001}{Paper~II}
\defcitealias{Weigelt_etal2002}{Paper~III}

\begin{document}

\title
{
{\irc} in action: present episode of intense mass-loss
reconstructed by two-dimensional radiative transfer modeling
}

\author
{
A.\,B. Men'shchikov\and
K.-H. Hofmann\and
G. Weigelt
}

\institute
{
Max-Planck-Institut f{\"u}r Radioastronomie, Auf dem H{\"u}gel 69,
D--53121 Bonn, Germany\\
e-mails: sasha, khh, weigelt (@mpifr-bonn.mpg.de)
}

\offprints{A.\,B.\,Men'shchikov}

\date{Received / Accepted}

\titlerunning{{\irc} in action: episode of intense mass-loss}
\authorrunning{Men'shchikov et al.}


\abstract
{
We present two-dimensional (2D) radiative transfer modeling of {\irc} at
selected moments of its evolution in 1995--2001, which correspond to three
epochs of our series of 8 near-infrared speckle images \citep{Osterbart_etal2000,
Weigelt_etal2002}. The high-resolution images obtained over the last 5.4 years
revealed the dynamic evolution of the subarcsecond dusty environment of {\irc}
and our recent time-independent 2D radiative transfer modeling reconstructed
its physical properties at the single epoch of January 1997
\citep{Men'shchikov_etal2001}. Having documented the complex changes in the
innermost bipolar shell of the carbon star, we incorporate the evolutionary
constraints into our new modeling to understand the physical reasons for the
observed changes. The new calculations show that our previous static model is
consistent with the brightness variations seen in the near-infrared images,
implying that during the last 50 years, we have been witnessing an episode of a
steadily increasing mass loss from the central star, from $\dot{M} \approx
10^{-5}$ {\Msun}\,yr$^{-1}$ to the rate of $\dot{M} \approx 3 \times 10^{-4}$
{\Msun}\,yr$^{-1}$ in 2001. The rapid increase of the mass loss of {\irc} and
continuing time-dependent dust formation and destruction caused the observed
displacement of the initially faint components C and D and of the bright cavity
A from the star which has almost disappeared in our images in 2001. Increasing
dust optical depths are causing strong backwarming that leads to higher
temperatures in the dust formation zone, displacing the latter outward with a
velocity $v_T \approx 27$ km\,s$^{-1}$ due to the evaporation of the recently
formed dust grains. This self-regulating shift of the dust density peak in the
bipolar shell mimics a rapid radial expansion, whereas the actual outflow has
probably a lower speed $v < v_\infty \approx 15$ km\,s$^{-1}$. The model
predicts that the star will remain obscured until $\dot{M}$ starts to drop back
to lower values in the dust formation zone; in a few years from that moment, we
could be witnessing the star reappearing.
\keywords
{
radiative transfer --
circumstellar matter --
stars: individual: {\irc} --
stars: mass-loss --
stars: AGB and post-AGB --
infrared: stars
}
}
\maketitle


\section{Introduction}
\label{Introduction}

The pulsating carbon star {\irc} (also known as \object{CW Leo}, \object{AFGL
1381}), together with its huge circumstellar envelope lost during its long
evolution on the asymptotic giant branch (AGB), is the best studied object of
its class. Being in a very advanced phase of its life, probably in transition
to protoplanetary nebulae \citep[][ hereafter Paper~I]{Osterbart_etal2000},
{\irc} presently exhibits a very high mass-loss rate $\dot{M} \sim 10^{-4}$
{\Msun}\,yr$^{-1}$ \citep[][ hereafter Paper~II]{Men'shchikov_etal2001}. After
three decades of intensive observational and theoretical work \citepalias[see,
e.g., references in][]{Men'shchikov_etal2001}, recent near-infrared speckle
imaging has revealed an extremely complex evolution of its circumstellar
material in the vicinity of the dust condensation zone on a time scale of one
year.

Near-infrared images with resolutions better than 100 mas presented by
\citep{Weigelt_etal1997,Osterbart_etal1997,Weigelt_etal1998a,Weigelt_etal1998b,
HaniffBuscher1998, Osterbart_etal2000,Tuthill_etal2000,Weigelt_etal2002} have
demonstrated that the inner, subarcsecond dust shell of {\irc} is non-spherical
and clumpy, with four components A, B, C, and D clearly visible. The detailed
two-dimensional radiative transfer modeling presented in
\citetalias{Men'shchikov_etal2001} has shown that the star is actually located
at the position of the second brightest component B. The brightest southern
peak A was identified with the radiation emitted and scattered in the optically
thinner cavity of the dense circumstellar shell. The model reconstructed
physical properties of the star and dusty envelope of {\irc} at a single moment
corresponding to the epoch of our high-resolution $H$- and $K$-band images on
January 23, 1997 \citepalias{Osterbart_etal2000}. As much as it was possible,
the model took into account most other observations of dust radiation, although
stellar pulsations and non-periodic changes of the shell made many measurements
from various epochs fundamentally incomparable in the frame of the static
model.

In the present study, we attempt to attack the problem using a simplified
approach based on the self-consistent model for a single epoch that we have
constructed in \citetalias{Men'shchikov_etal2001}. The idea is to extend this
modeling to the first and to the last epochs of the 6-year sequence of
high-resolution $K$-band speckle images we have obtained since October 1995
\citep[][ hereafter Paper~III]{Weigelt_etal2002}. In the beginning of our
monitoring of {\irc}, component B (the star) was relatively bright. The direct
light from the star has been gradually fading since then, whereas the bright
lobe A (the southern cavity) has become dominant. The angular distance between
components A and B increased from $\sim$ 190 to $\sim$ 350 mas between 1995
and 2001, implying a linear speed $v_{\rm A} \approx 18$\,km\,s$^{-1}$ in the
plane of sky and a deprojected radial velocity $v_{r,{\rm A}} \approx
19$\,km\,s$^{-1}$ \citepalias[Appendix A of][]{Men'shchikov_etal2001}, for
the assumed distance of $D = 130$ pc and the viewing angle of $\theta_{\rm v} =
40${\degr} \citepalias{Men'shchikov_etal2001}. On the basis of our
time-independent model, this well-documented rapid evolution has been
qualitatively interpreted in terms of dust formation in a bipolar stellar wind,
the increasing mass-loss rate, and the sublimation of the recently formed
grains in a progressively hotter region just outside the dust formation radius
\citepalias{Weigelt_etal2002}.

The goal of the present study is to determine, whether the observed changes in
the high-resolution speckle images are consistent with our previous model of
{\irc}, and to derive more accurate physical parameters of the wind at several
moments in time. In Sect.~\ref{RadTraModel}, we describe our approach and model
parameters. In Sect.~\ref{Results}, we discuss the results of our new modeling,
comparing them with the high-resolution images, intensity profiles, and
spectral energy distribution (SED) of {\irc}. In Sect.~\ref{Conclusions}, we
summarize the model parameters and our conclusions.


\section{Radiative transfer model}
\label{RadTraModel}


\subsection{General formulation}
\label{General}

Our new 2D radiative transfer modeling is based on the structure and physical
properties of {\irc} derived in \citetalias{Men'shchikov_etal2001}. We refer to
the latter for a detailed description of the model geometry and discussion of
all assumptions. The parameters of the envelope were kept unchanged, except for
the density distribution within 100 AU from the star and the cavity's opening
angle (see Sects.~\ref{Parameters} and \ref{Images}). The density profile of
the circumstellar shell on subarcsecond scales was allowed to vary due to
expansion, increased mass-loss rate, and condensation of new dust in the wind
(Sects.~\ref{Images}, \ref{DensTem}, and \ref{MassLoss}).

The new models were computed for selected epochs (1995, 1997, and 2001)
spanning the entire sequence of our speckle images in the $K$ band
\citepalias{Weigelt_etal2002}. Slightly adjusting the density distribution of
the time-independent model of \citetalias{Men'shchikov_etal2001}, we found the
models which would produce the images and intensity profiles consistent with
the observed ones. We limited ourselves to the constraints from the
high-resolution near-infrared images and did not attempt to perform a new
global search of the entire parameter space for each epoch in order to fit all
observations. This approach is justified by the fact that the latter have
already been explained self-consistently by our previous model. Moreover, it is
highly problematic to use a mixture of old observational constraints in a
meaningful way for the rapidly evolving inner shell of {\irc}.

\begin{figure*}
\begin{center}
\resizebox{0.75\hsize}{!}{\hspace{1mm}}
\vspace{0.3mm}
\resizebox{0.75\hsize}{!}
   {
    \includegraphics{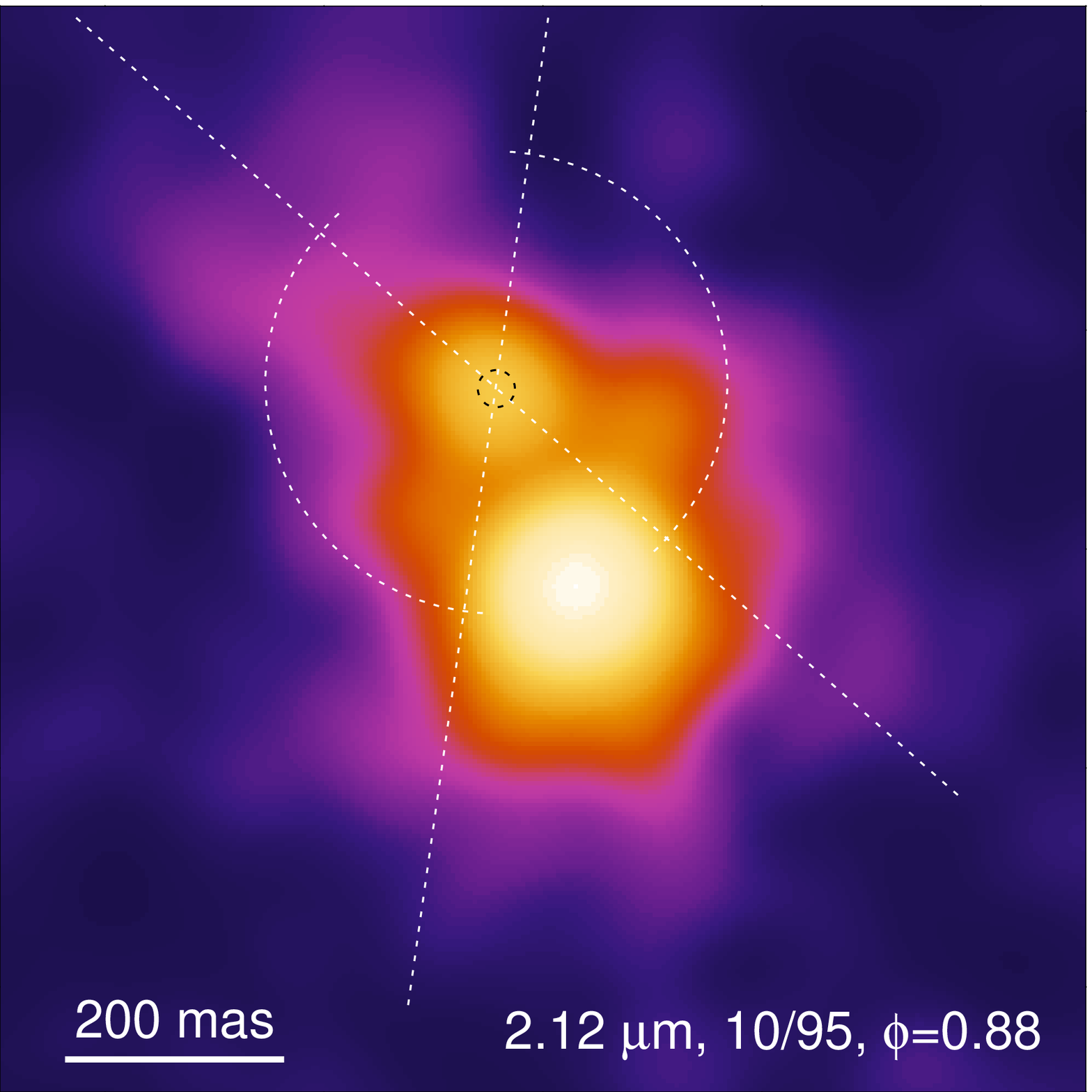}
    \hspace{0mm}
    \includegraphics{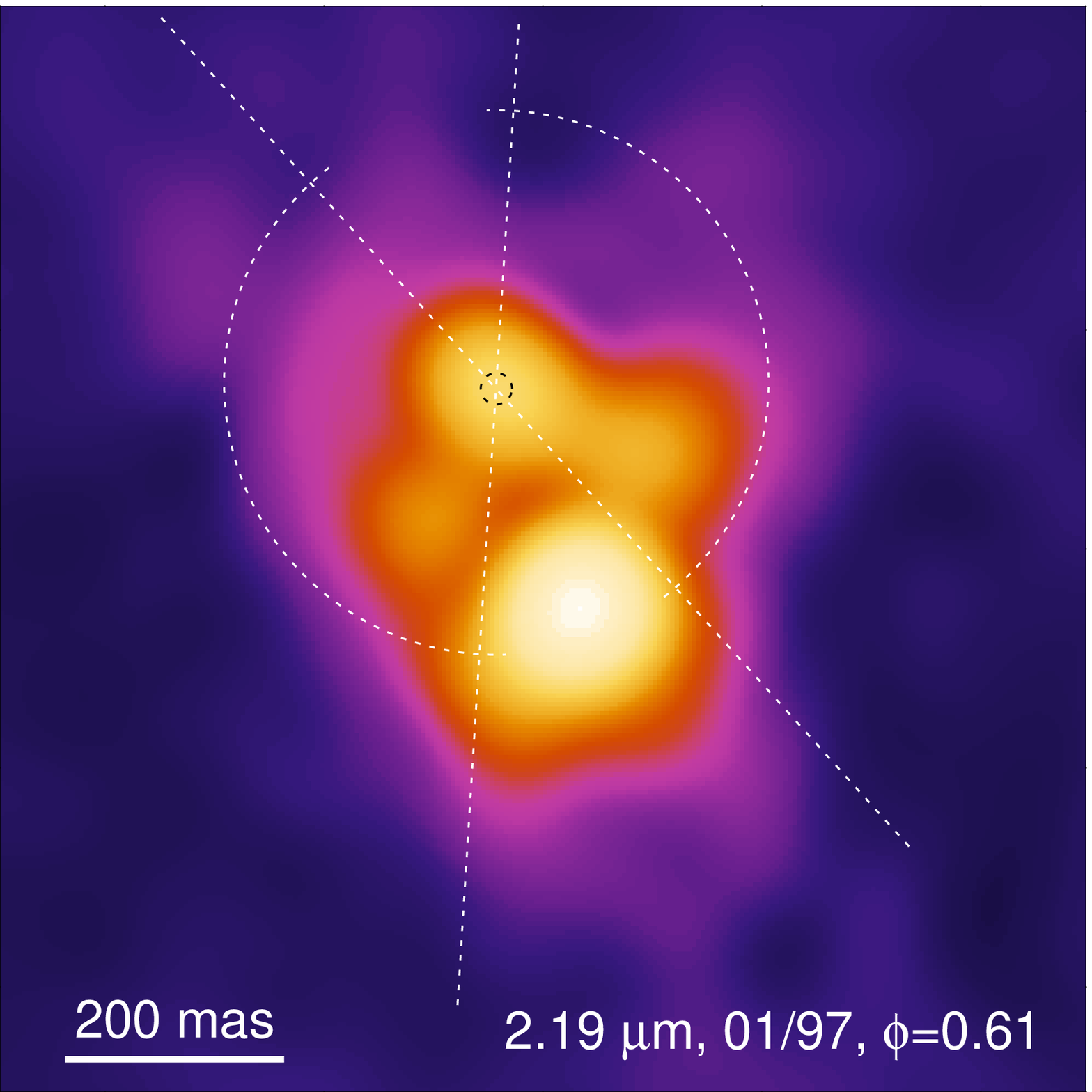}
    \hspace{0mm}
    \includegraphics{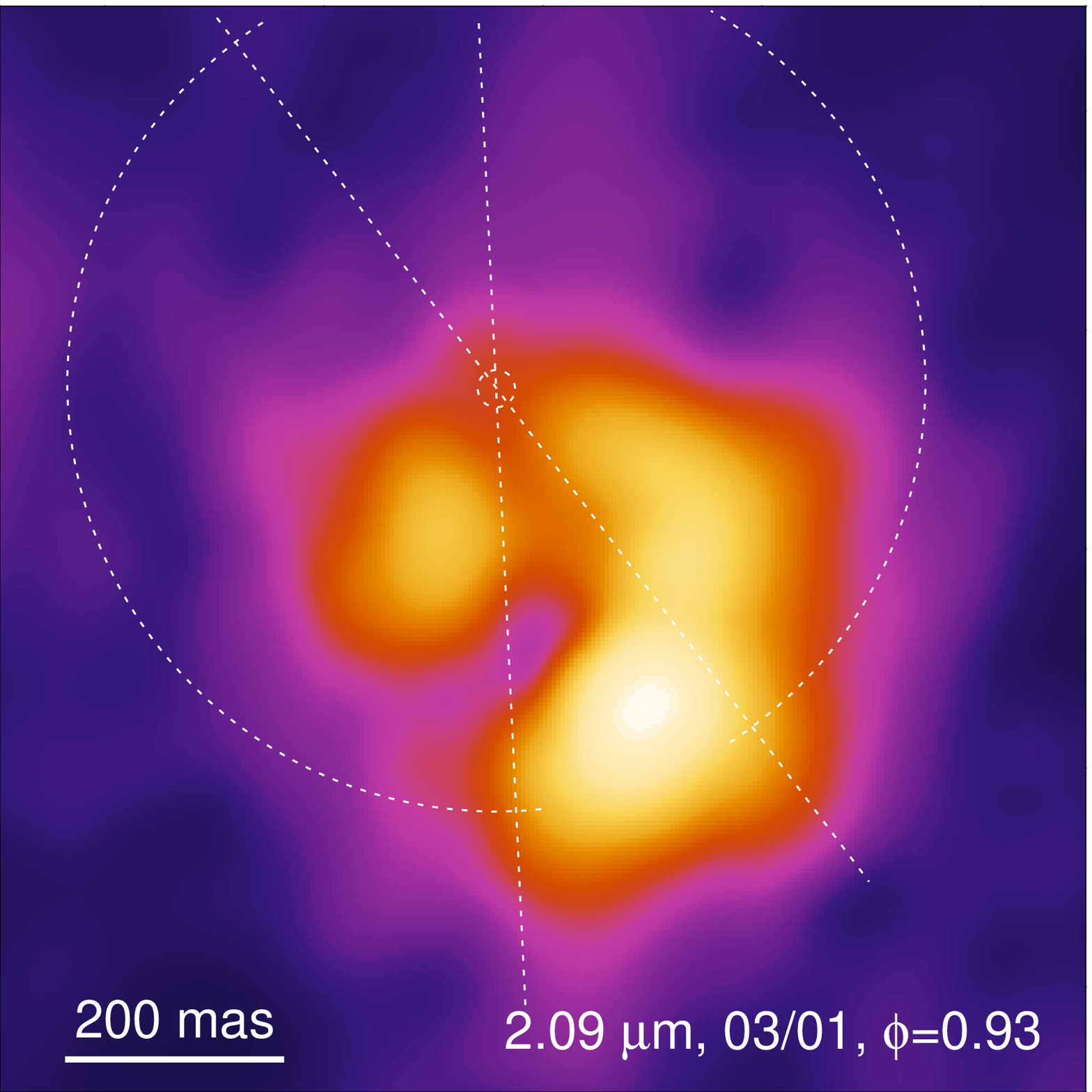}
   }
\resizebox{0.75\hsize}{!}
   {
    \includegraphics{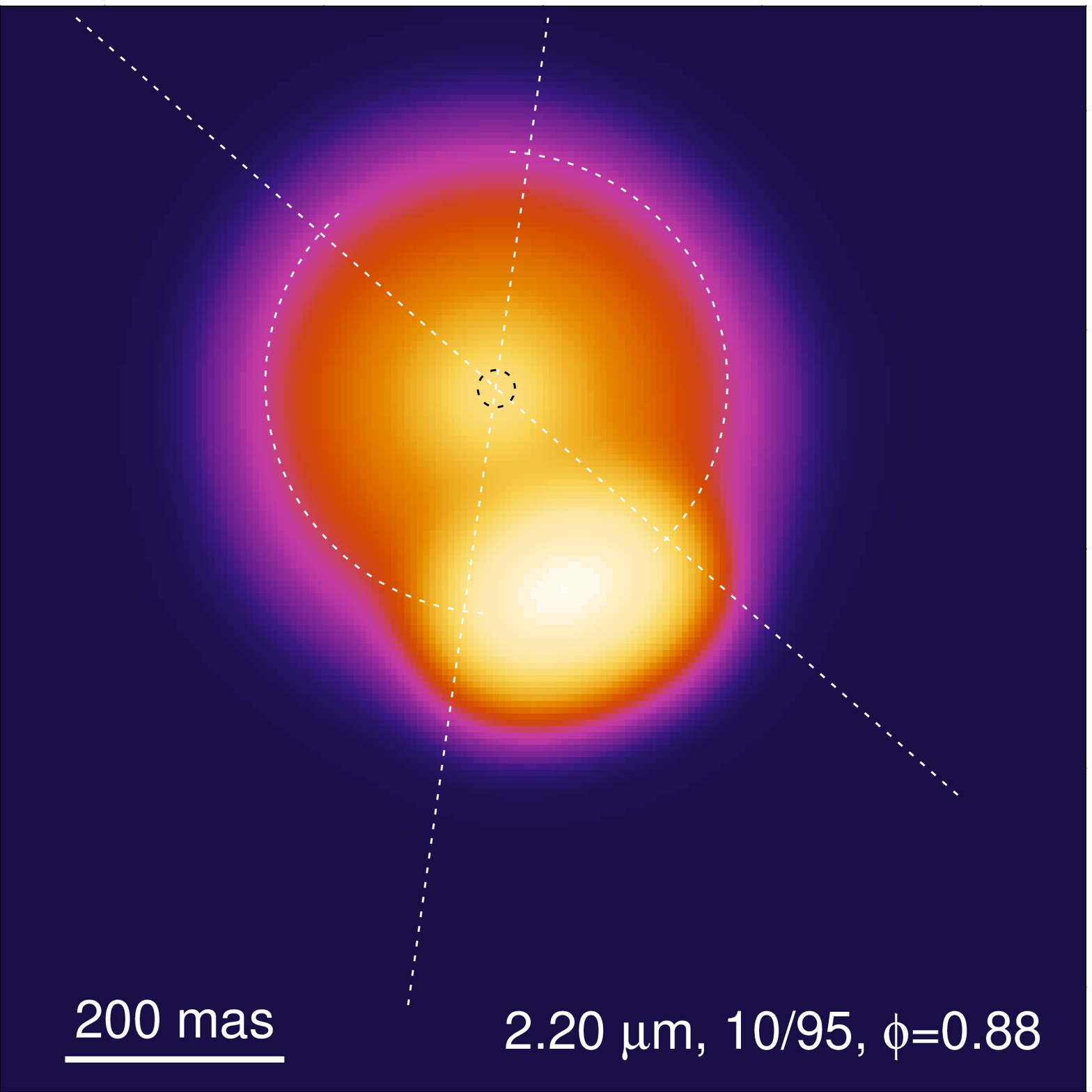}
    \hspace{0mm}
    \includegraphics{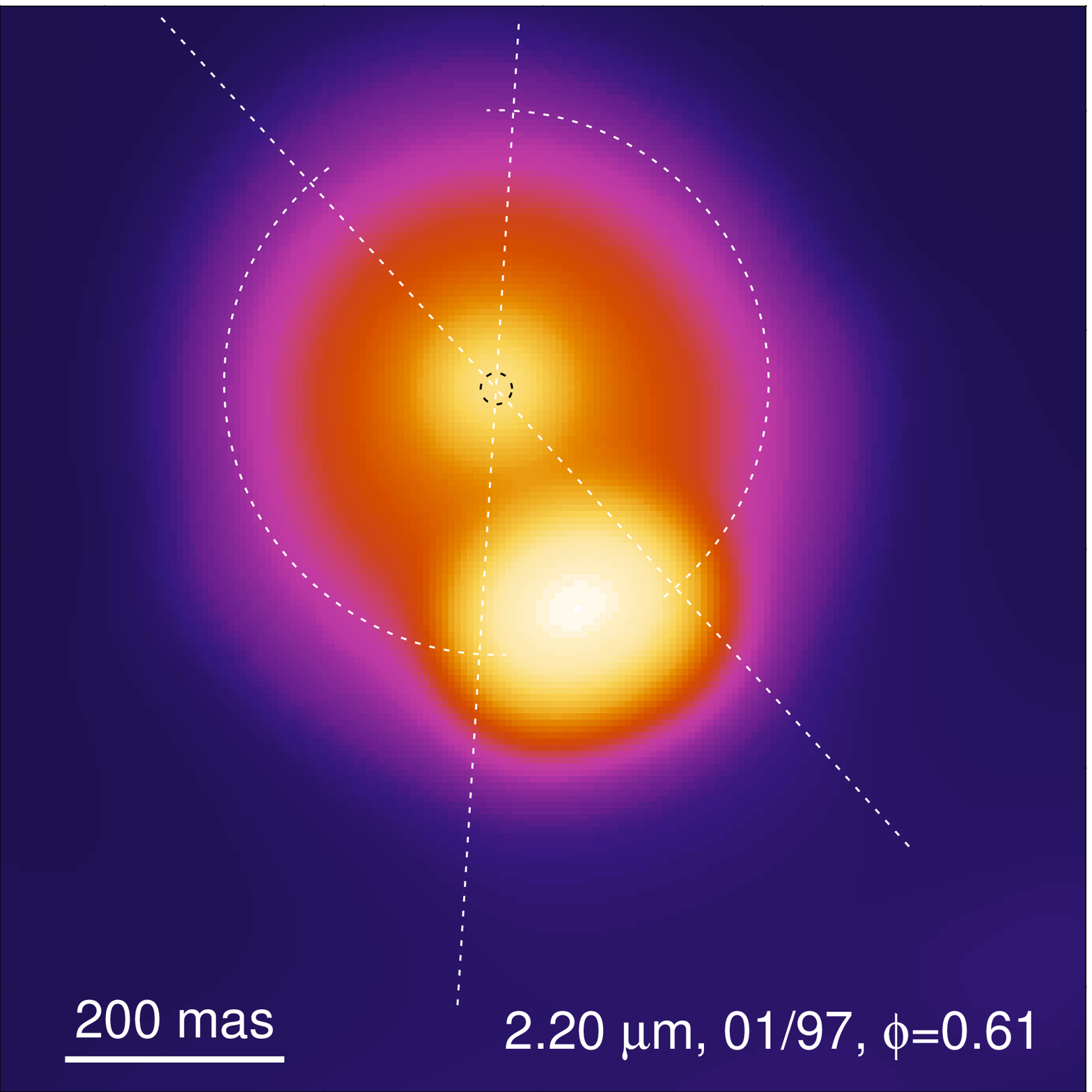}
    \hspace{0mm}
    \includegraphics{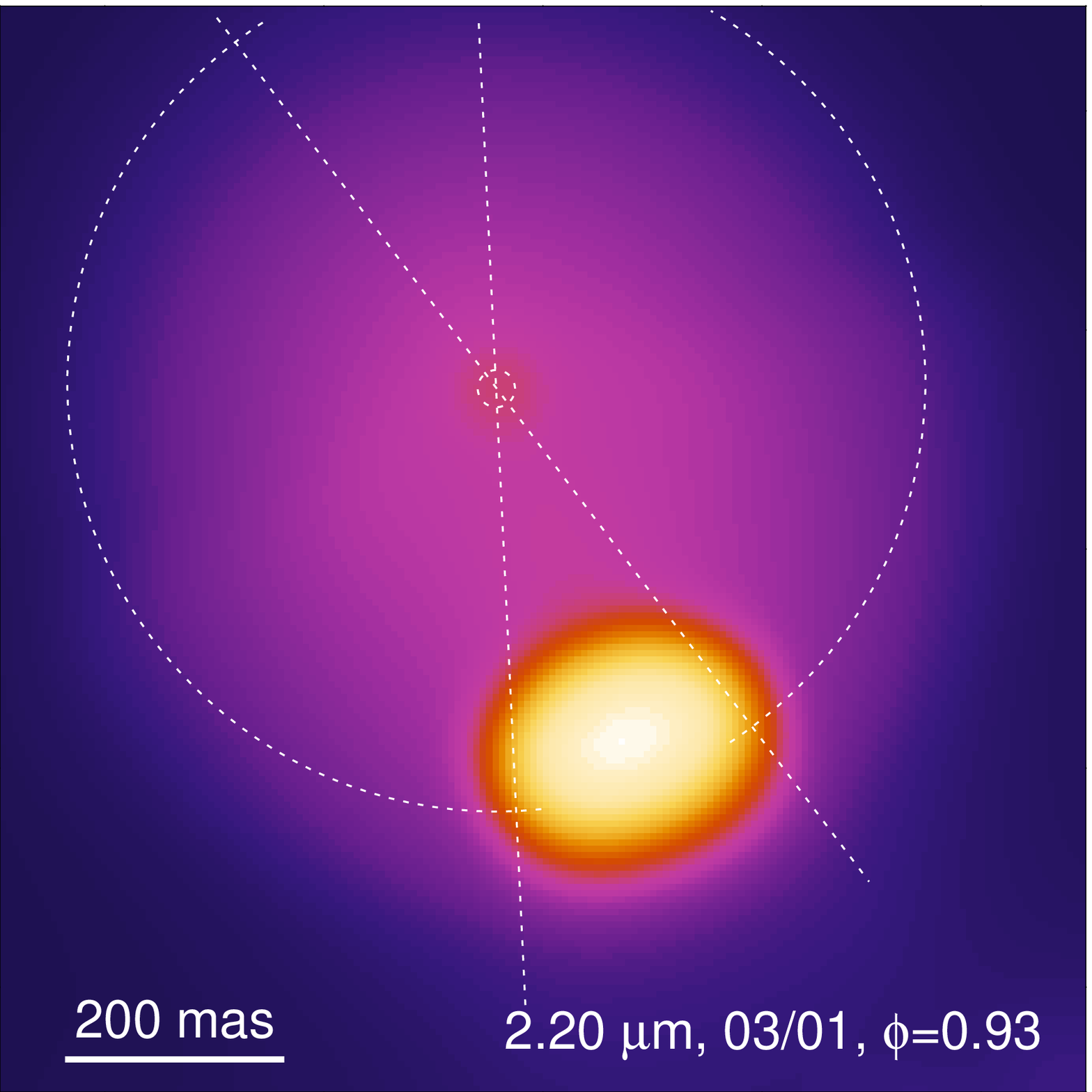}
   }
\caption
{
Comparison of our new $K$-band model images ({\em lower panels}) with our
$K$-band speckle images of {\irc} ({\em upper panels}) obtained in October
1995, January 1997, and March 2001 with resolutions of 92, 87, and 73\,mas,
respectively \citepalias{Weigelt_etal2002}. The two dashed circular segments
show the radius of the density peak in the main dust formation zone of our
model (Fig.~\ref{DenTem}) and the small dashed circles indicate the position
and actual size of the star. The straight dashed lines intersecting at the
star's center outline the conical bipolar cavities of the model. Each image
shows a 1{\arcsec}$\,\times\,$1{\arcsec} area. North is up and east is to the
left.
}
\label{Kimages}
\end{center}
\end{figure*}
\begin{figure*}
\begin{center}
\resizebox{0.75\hsize}{!}{\hspace{1mm}}
\vspace{0.3mm}
\resizebox{0.75\hsize}{!}
   {
    \includegraphics{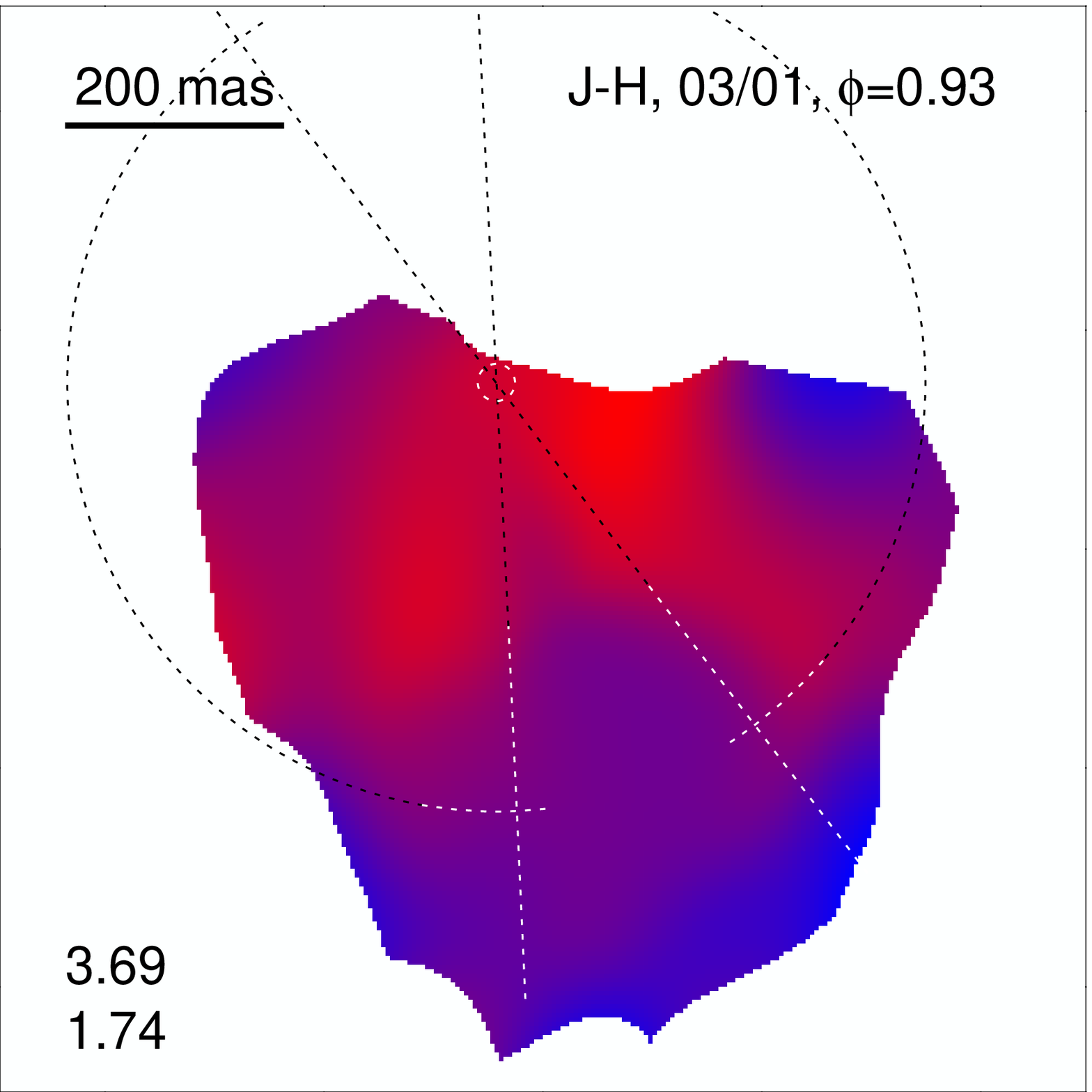}
    \hspace{0mm}
    \includegraphics{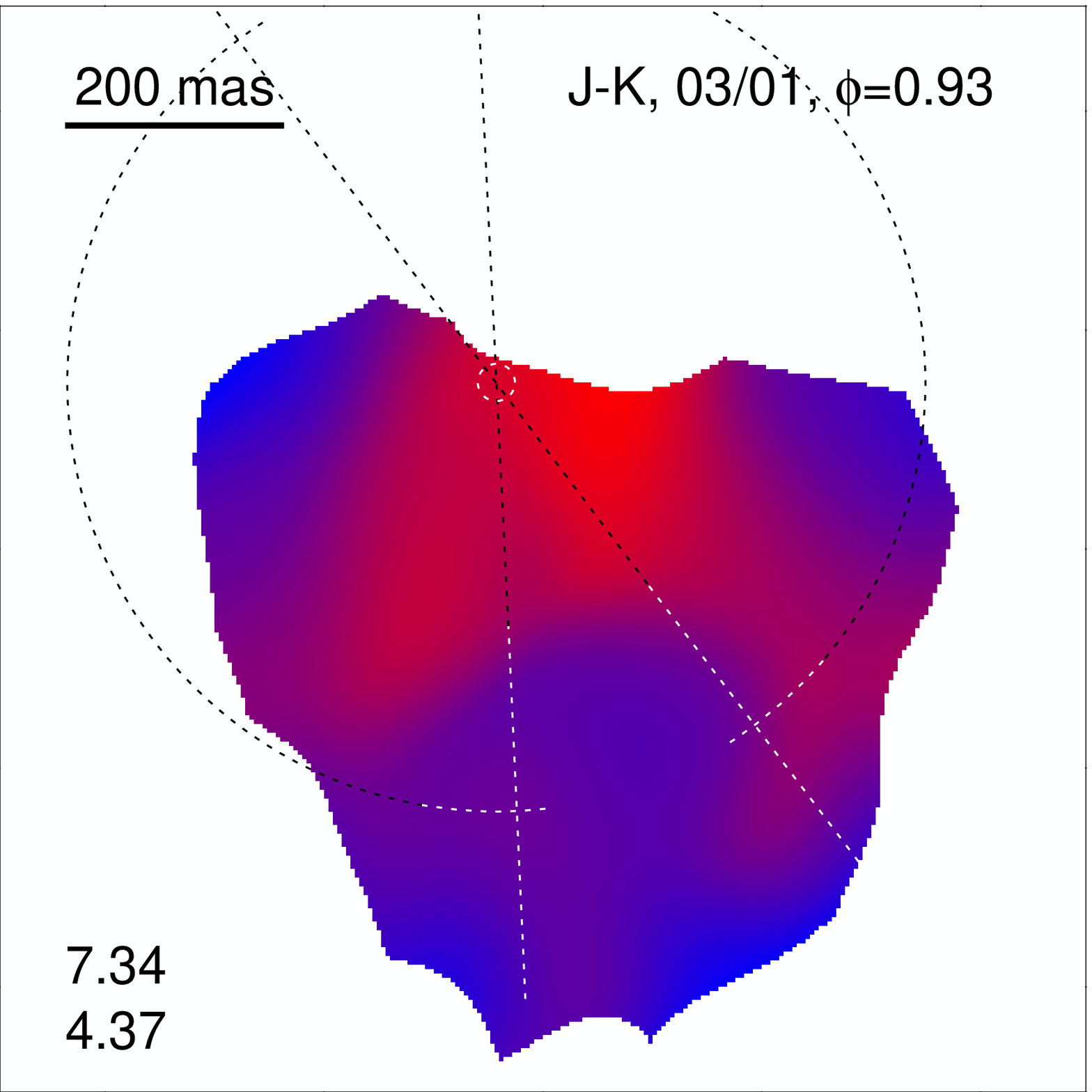}
    \hspace{0mm}
    \includegraphics{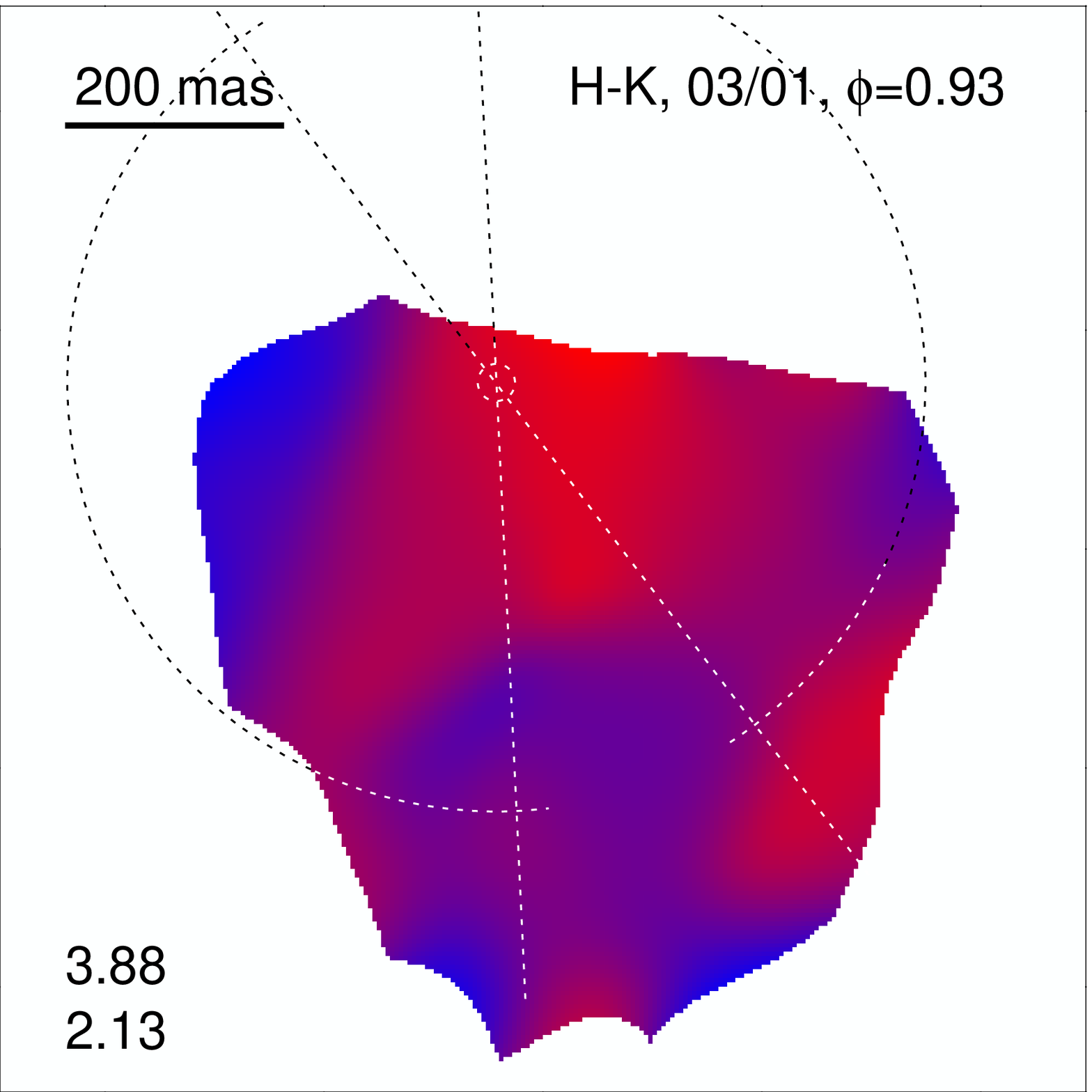}
   }
\resizebox{0.75\hsize}{!}
   {
    \includegraphics{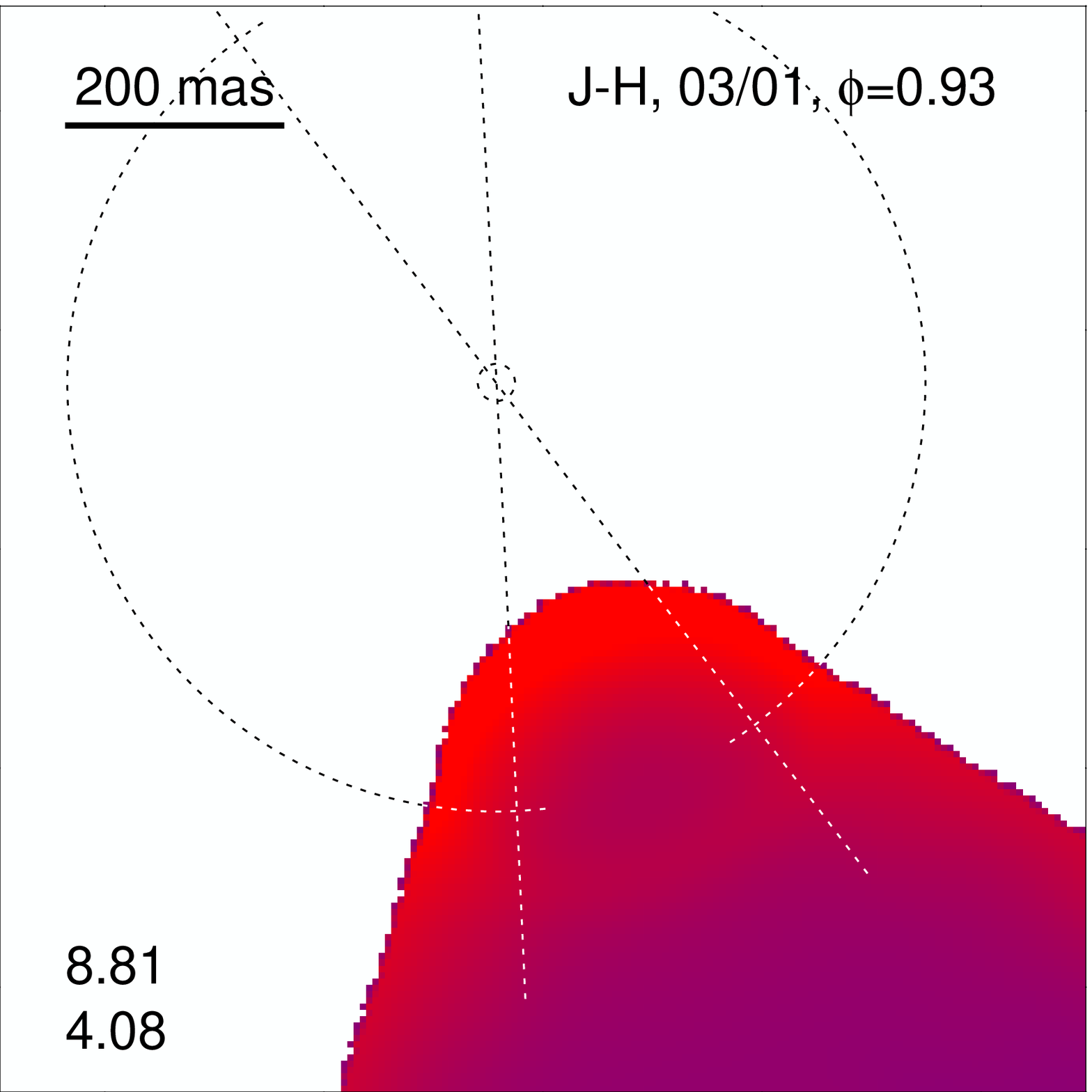}
    \hspace{0mm}
    \includegraphics{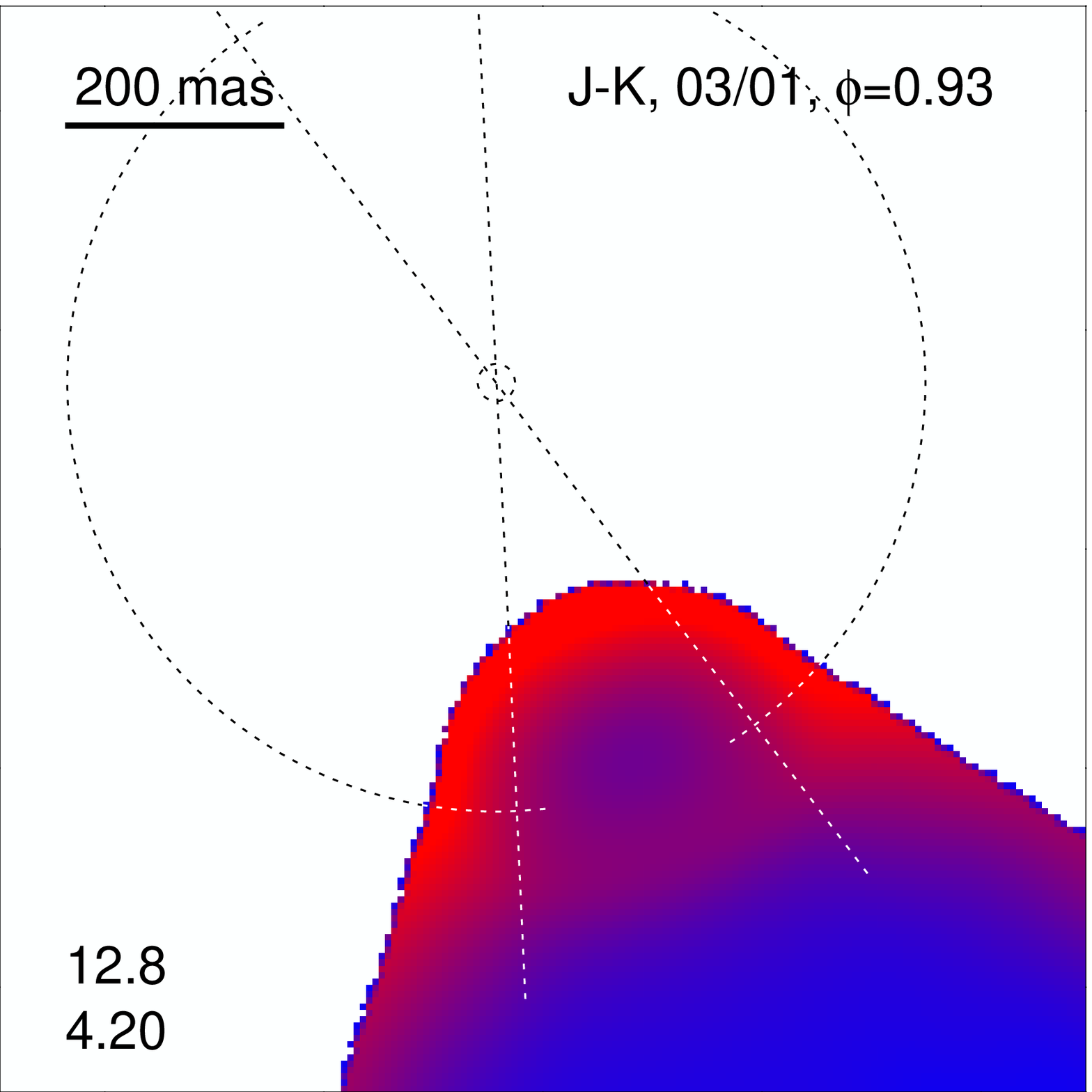}
    \hspace{0mm}
    \includegraphics{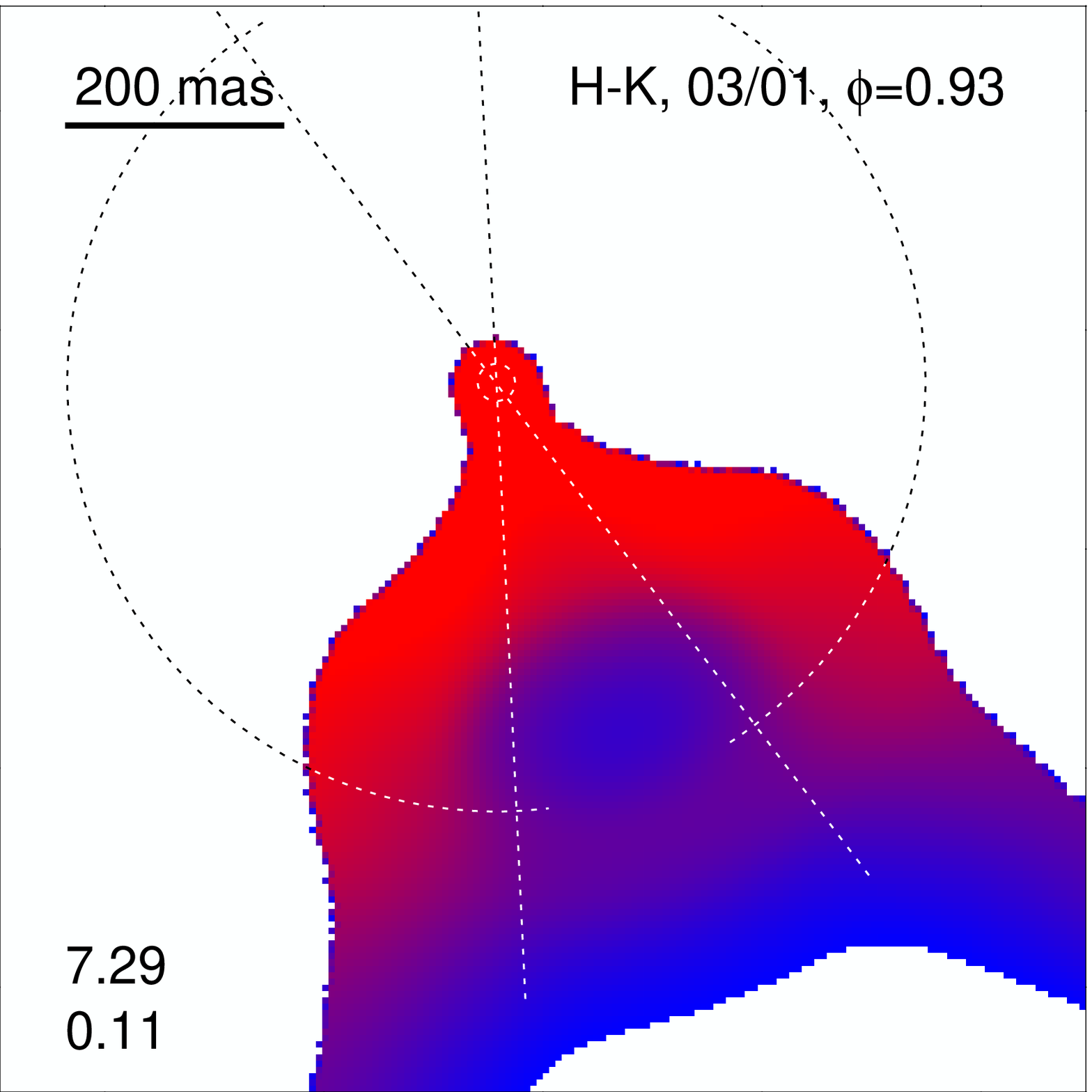}
   }
\caption
{
Comparison of the {\JmH}, {\JmK}, and {\HmK} model color images ({\em lower
panels}) with our latest color images of {\irc} ({\em upper panels}) obtained
in March 2001 with a resolution of 73 mas \citepalias{Weigelt_etal2002}. The
observed colors were computed for only those pixels where the intensities in
$H$ and $K$ are higher than 2\,{\%} of the brightest peak, whereas the model
colors are displayed for the areas brighter that 1\,{\%}. The pairs of numbers
in the lower corners are the maximum and minimum colors. The
two dashed circular segments show the radius of the density peak in the main
dust formation zone of our model at this epoch and the small dashed circles
indicate the position and actual size of the star. The straight dashed lines
intersecting at the star's center outline the conical bipolar cavities of the
model. Each image shows a 1{\arcsec}$\,\times\,$1{\arcsec} area. North is up
and east is to the left.
}
\label{Colors}
\end{center}
\end{figure*}


\subsection{Model parameters}
\label{Parameters}

To facilitate the presentation and understanding of the new modeling, we
briefly describe the parameters of our previous, time-independent model of
{\irc}. At the end of this section, we summarize the time-dependent parameters
of the new models.

The study reported in \citetalias{Men'shchikov_etal2001} reconstructed the
following physical picture. For the adopted distance $D$ = 130 pc, the
luminosity {\Lstar} of the pulsating star varies from 13000 to 5200 {\Lsun}
between the maximum and minimum brightness, with a period of 649 days. The
effective temperature {\Tstar} of the star is changing from 2800 to 2500 K
between the two phases, whereas the stellar radius {\Rstar} varies in a
relatively narrow range, from 500 to 390 {\Rsun}. During the last 1000 years,
the star has experienced at least two episodes of very high mass loss with
rates reaching $\dot{M} \sim 10^{-4}$ {\Msun}\,yr$^{-1}$. The
broken-power-law density profile of our model \citepalias[see Fig.~\ref{DenTem};
also Table~6 in][]{Men'shchikov_etal2001} results most likely from a complex
interplay between stellar pulsations, changing mass-loss rate, and radiative
acceleration of dust and gas in the innermost parts of the envelope. The latter
has bipolar cavities with a full opening angle $\omega = 36${\degr}, tilted
toward us by $\theta_{\rm v} = 40${\degr} from the sky plane. The huge envelope
of a radius $R_2 = 3$ pc has a total mass $M = 3$ {\Msun}, which corresponds to
a dust-to-gas mass ratio {\dustgas} = 0.0039. The mass is consistent with an
initial mass of the central star {\Mstaro} $\approx 4$ {\Msun} and a
present-day core mass {\Mstar} $\approx 0.7$ {\Msun}. Different dust components
exist in the envelope of {\irc} everywhere from the stellar photosphere to the
outer boundary, producing the total visual optical depth ${\tau}_V \approx 40$
toward the star, whereas in the polar regions it is ${\tau}_{{\rm p}V} \approx
10$.

Both the radial distribution and composition of dust in the model of {\irc}
were strongly inhomogeneous. The innermost dust formation zone of the envelope
was located very close to the stellar photosphere, where the newly formed SiC
grains had temperatures $T_{\rm SiC} \approx 2000$ K. Most of the carbonaceous
dust nucleated further away from the star (at $r_{\rm C} = 30$ AU at that
epoch), whereas the most volatile components condensed at even larger distances
(at $r_{\rm [Mg,Fe]S} = 45$ AU). The abundances and size distribution of dust
grains in our model depend on the distance from the star. The modeling
identified the carrier of the 11.3 {\mic} feature with {\SiCC} particles, i.e.
unorganized aggregates made of incompletely amorphous carbon grains with
significant graphitic content and silicon carbide grains. The 27 {\mic}
emission band, also known as the 30 {\mic} feature, was well reproduced by the
core-mantle {\SiCCMgFeS} grains.

The new models of {\irc} presented in this paper correspond to the epochs of
October 8, 1995, January 23, 1997, and March 9, 2001 (below, we refer to
the epochs only by the years). We fixed the model parameters at the values
derived in our previous model except for only the star's properties and the
envelope's density distribution \citepalias{Men'shchikov_etal2001}. The latter
resulted from the most recent episode of increased mass loss (within a radius
of $\sim 100$ AU, Sect.~\ref{DensTem}). Thus, our model assumes that at the
first epoch of our imaging series, in October 1995, the star had {\Lstar}$_1$ =
12200 {\Lsun}, {\Tstar}$_1$ = 2760 K, and {\Rstar}$_1$ = 483 {\Rsun} (17.3
mas); at the third epoch, in January 1997, {\Lstar}$_3$ = 6140 {\Lsun},
{\Tstar}$_3$ = 2530 K, and {\Rstar}$_3$ = 408 {\Rsun} (14.6 mas); and at the
eighth epoch, in March 2001, {\Lstar}$_8$ = 12900 {\Lsun}, {\Tstar}$_8$ = 2790
K, and {\Rstar}$_8$ = 486 {\Rsun} (17.4 mas). The cavities' opening angle
changed from $\omega_1 = 46${\degr} to $\omega_3 = 36${\degr} to $\omega_8 =
26${\degr} between the epochs (Sect.~\ref{CavityShapes}). The latter correspond
to the luminosity phases $\phi_1 = 0.88$, $\phi_3 = 0.61$, and $\phi_8 = 0.93$,
respectively. The intervals between the epochs amount to $\Delta t_{13} =
473$ days and $\Delta t_{38} = 1506$ days.


\section{Results and discussion}
\label{Results}


\subsection{Model images and intensity profiles}
\label{Images}

\begin{figure}
\begin{center}
\resizebox{0.858\hsize}{!}
   {
    \includegraphics{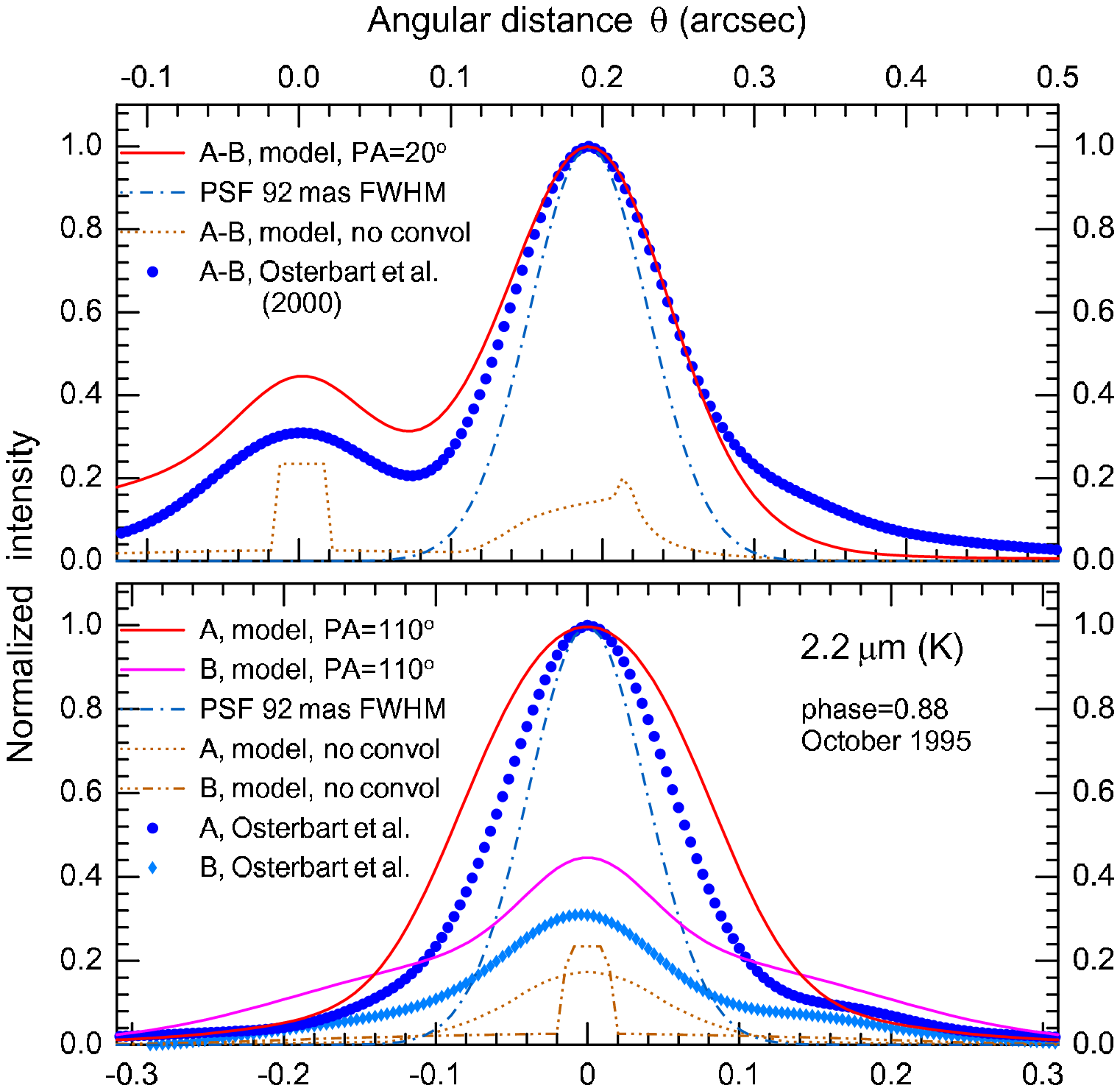}
   }
\resizebox{0.858\hsize}{!}
   {
    \includegraphics{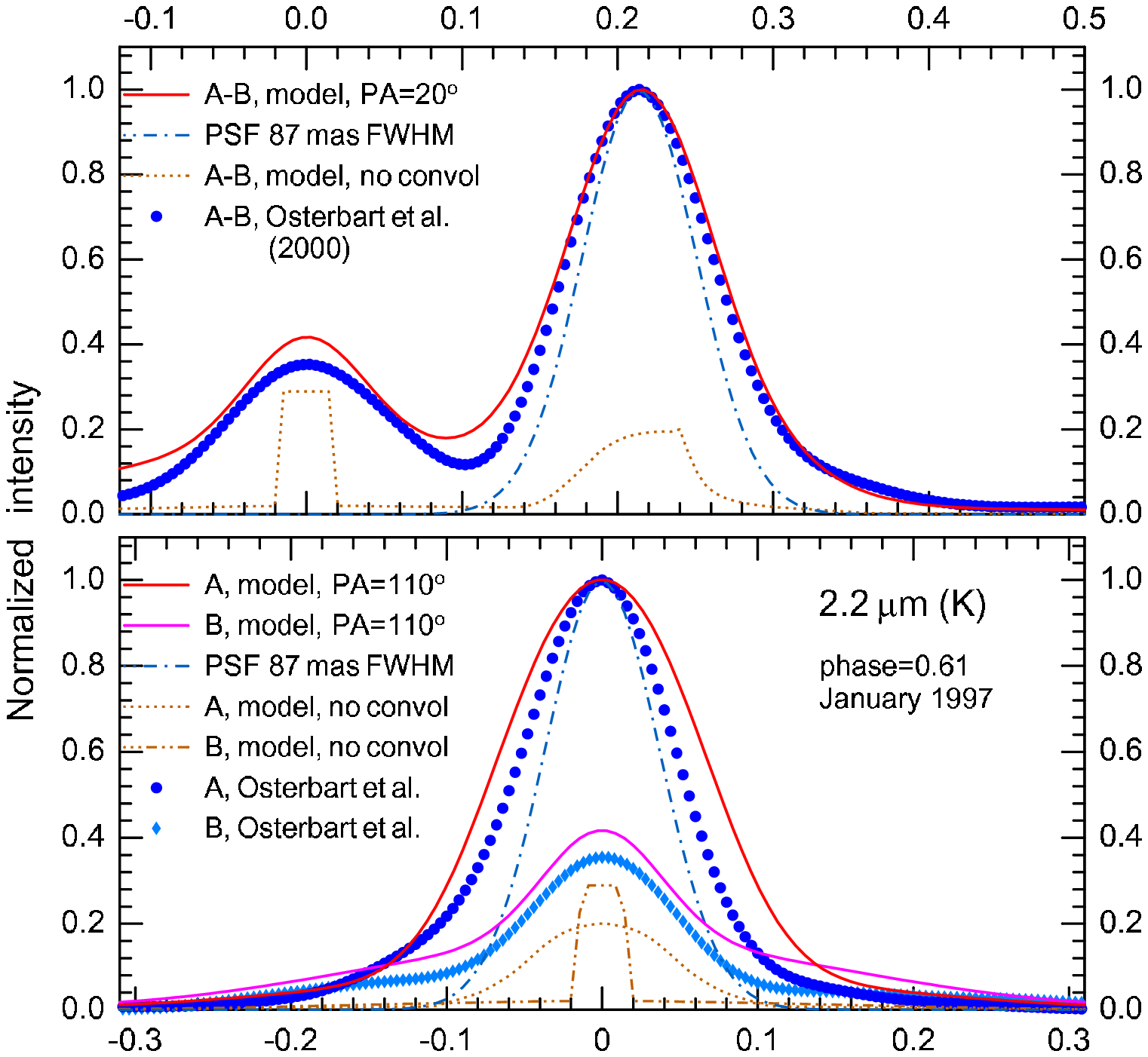}
   }
\resizebox{0.858\hsize}{!}
   {
    \includegraphics{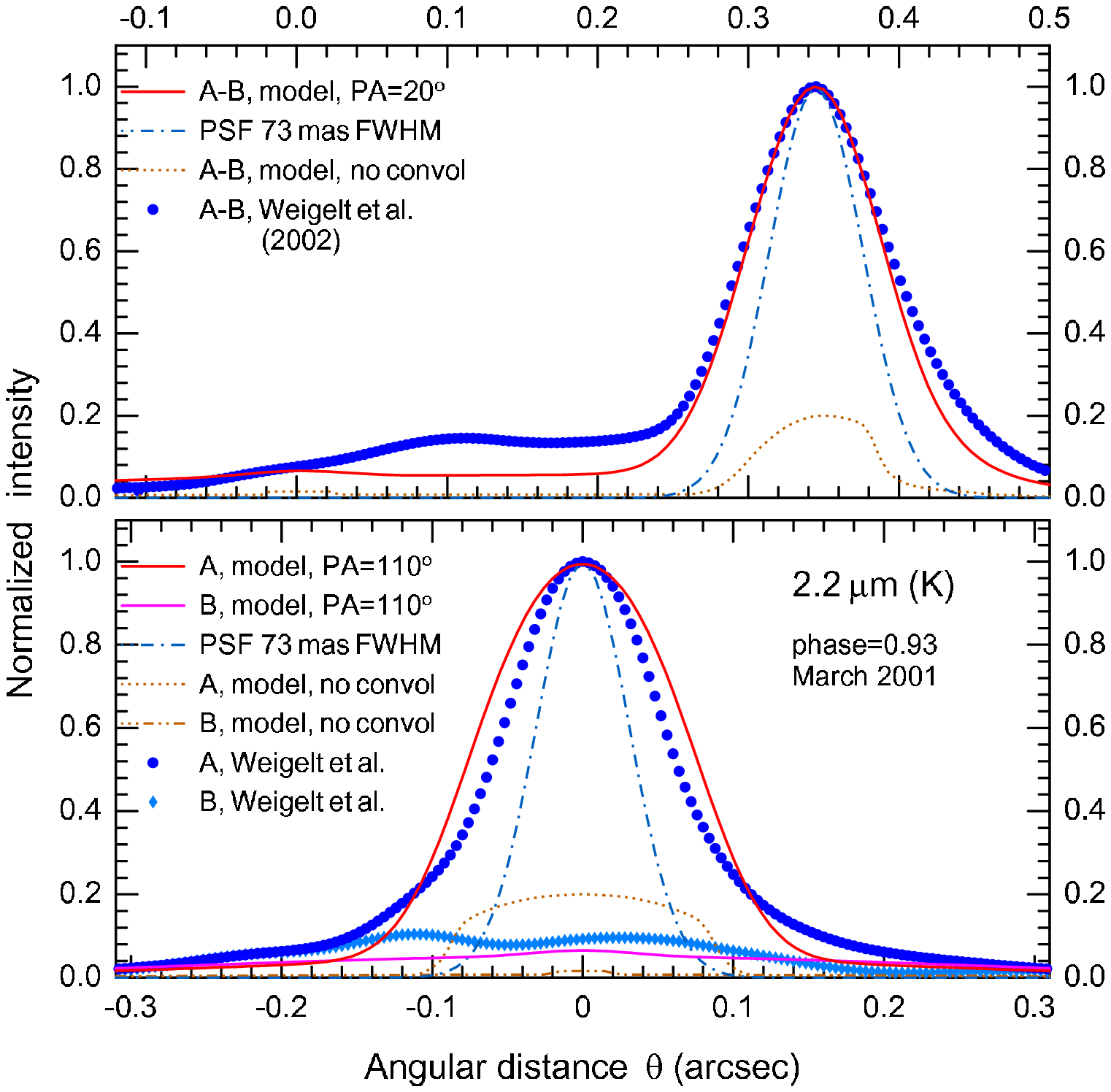}
   }
\caption
{
$K$-band model intensity profiles compared with the intensity cuts (in two
orthogonal directions) extracted from our images of {\irc} obtained in 1995,
1997, and 2001. The unconvolved model intensity cuts through the star and the
southern cavity show the ``true'' model intensity distributions with a
numerical resolution of 5 mas.
}
\label{Knorm}
\end{center}
\end{figure}

Figure~\ref{Kimages} compares our new $K$-band model images with the
high-resolution speckle images of {\irc} obtained in 1995, 1997, and 2001.
Figure~\ref{Colors} displays the {\HmK}, {\JmH}, and {\JmK} model color maps
and the observed color distributions derived from the $J$, $H$, and $K$ speckle
images of 2001. Figure~\ref{Knorm} shows a more quantitative comparison of the
model and observed intensity distributions in terms of the normalized intensity
profiles. The latter were derived from the images in two directions, parallel
(PA $\approx 20${\degr}) and perpendicular (PA $\approx 110${\degr}) to the
projected symmetry axis of the bipolar structure. We define the axis as a
straight line through the peaks A and B in the $K$ image of 1995, where the
peaks are best visible. Before extracting the profiles, we convolved the model
intensity distributions with the point-spread functions (PSF) of 93, 87, and 73
mas full width at half-maximum (FWHM) corresponding to the resolutions of the
observed images.

To facilitate direct comparisons with observations, an outline of the model
geometry is displayed in Figs.~\ref{Kimages} and \ref{Colors} by dashed lines.
The central star is shown by a small dashed circle whose
diameter corresponds to the actual angular size of the star at the epochs
(Sect.~\ref{Parameters}). The straight lines intersecting at the star's center
indicate the biconical geometry of the cavities with a projected opening angle
$\hat{\omega}$. The cones' axis is tilted by $\theta_{\rm v} = 40${\degr} with
its southern side toward the observer, causing the apparently increased
cavities' opening from $\omega$ to $\hat{\omega} = 2
\arctan\,(\tan\,({\omega}/2)\,{\cos}^{-1}\theta_{\rm v})$. Two circular
segments extended over the angle $\pi - \omega$ are placed in each image at the
density peak position in the main dust formation zone of the model. The
segments' radii increased with time from $r_1 = 28$ AU to $r_3 = 33$ AU to $r_8
= 55$ AU (Sect.~\ref{DensTem}), reflecting the observed displacement of
component A from the star.

\subsubsection{Motion of the components}
\label{Motion}

If the bright peaks in our near-IR images of {\irc} were clumps of gas and
dust, their increasing distance from the star could be interpreted as real
motion of the local density enhancements. Our radiative transfer modeling in
\citetalias{Men'shchikov_etal2001} has shown, however, that the brightest peak
A seen in all images is not the direct light from the central star. The star is
actually located at the position of the fainter peak B, whereas component A is
the radiation emitted and scattered in the optically thinner polar cavity of
the optically thick dust shell. The even fainter components C and D were
identified with smaller-scale deviations of the density distribution of the
circumstellar environment from axial symmetry. Along the directions to these
peaks from the central star, circumstellar material is less opaque than in
other regions of the dense shell. We interpreted a difference in the {\em
apparent} velocities of the components ($v_{\rm A} \approx 14$ km\,s$^{-1}$ vs.
$v_{\rm C,D} \approx 5$ km\,s$^{-1}$) as a pure projection effect and
deprojected their common radial velocity of $v_r \approx 15$ km\,s$^{-1}$ for
1995--1997. New speckle images presented in \citetalias{Weigelt_etal2002}
revealed a significantly faster motion of component A in 1997--2001, with
$v_{\rm A} \approx 21$ km\,s$^{-1}$.

The apparent outward motion can be interpreted either as a {\em real radial
expansion} of the opaque, dense (spherical) layer with several optically
thinner cavities or as a {\em displacement of the dust formation radius} due to
the evaporation of recently formed dust in a hotter environment. Increasing
temperatures would evaporate grains and thus affect the location of the inner
dust boundary of the envelope, causing its displacement with a velocity $v_T =
[\partial T (r,t) / \partial t] \,[\partial T (r,t) / \partial r]^{-1}$.

Expansion of a dense layer with time-independent parameters of dust grains
would be inconsistent with the star clearly fading since 1997. In fact, any
formation and expansion of the detached shell would mean that a high mass-loss
rate has eventually dropped and that the dense layer leaves behind
lower-density material, as it moves outward. Since densities and optical depths
of such a shell would continuously decrease in the process of expansion, the
latter can only be associated with brightening, not fading of the star, unless
dust grains are still forming in the expanding layer, increasing its optical
depth. This does not seem plausible, because dust formation would quickly
become inefficient with decreasing gas densities.

A much better explanation for the decreasing brightness of the star is to
assume a monotonically rising mass-loss rate and, hence, higher densities and
optical depths of the wind in the vicinity of the dust formation radius. Note,
however, that the increasing $\dot{M}$ would not form a detached dense
shell and it is only possible to displace the dust boundary if the recently
formed dust grains are destroyed. This would also lead to decreasing optical
depths of the envelope and brightening of the star, unless dust grains continue
to form in the wind, making it increasingly opaque. This scenario is, however,
much more likely, since the gas density would peak in the dust formation zone
and it would be increasing with time.

To distinguish between the real motion of a dense layer and the temperature
shift of the dust formation zone, we need an accurate estimate of the apparent
velocities of components A, C, and D relative to the star. In fact, one can
expect an acceleration of the circumstellar material by the radiation pressure
on dust grains in the neighborhood of the dust formation zone, not its
deceleration. If the deprojected radial velocities were not higher than the
terminal outflow speed $v_{\infty} \approx 15$ km\,s$^{-1}$ observed in {\irc},
the apparent motion could reflect largely an expansion of the detached dense
layer in which grains are forming. The observed velocities are, however,
significantly higher than $v_{\infty}$.

For the assumed distance $D = 130$ pc, a linear fit to the separations of
component A \citepalias{Weigelt_etal2002} gives $v_{\rm A} \approx 18$
km\,s$^{-1}$ in the plane of sky, whereas a parabolic fit yields $v_{\rm A}
\approx 10$ km\,s$^{-1}$ in 1996 and suggests 26 km\,s$^{-1}$ in 2001
($\dot{v}_{\rm A} \approx$ 3 km\,s$^{-1}$\,yr$^{-1}$). Components C and D moved
with an average apparent velocity $v_{\rm C,D} \approx 5.5$ km\,s$^{-1}$ during
the years of our observations. Interpreting the differences of the components'
velocities in our model as pure projection effects \citepalias[cf. Appendix A
in][]{Men'shchikov_etal2001} and assuming that the acceleration of A is real,
we can expect for C and D an acceleration term $\dot{v}_{\rm C,D} \approx 1$
km\,s$^{-1}$\,yr$^{-1}$ ($v_{\rm C,D} \approx 3$ km\,s$^{-1}$ in 1996 and 8
km\,s$^{-1}$ in 2001). Within uncertainties of the measured separations, the
data presented in \citetalias{Weigelt_etal2002} are consistent with this
picture. For component A, one can deproject a radial velocity $v_{r{\rm A}}
\approx 19$ km\,s$^{-1}$ (or $v_{r{\rm A}} \approx 11$ km\,s$^{-1}$ in 1996 and
28 km\,s$^{-1}$ in 2001, for accelerated motion). For components C and D, the
velocity is $v_{r{\rm C,D}} \approx 17$ km\,s$^{-1}$ (or $v_{r{\rm C,D}}
\approx 9$ km\,s$^{-1}$ in 1996 and 27 km\,s$^{-1}$ in 2001, if there is
acceleration).

Since the acceleration is most likely real, the expansion velocity $v_r \approx
28$ km\,s$^{-1}$ is now twice higher than the terminal wind speed. Taken in
context with the increasing optical depths in the model shell ($\tau_{V1}
\approx 30$ in 1995, $\tau_{V3} \approx 37$ in 1997, and $\tau_{V8} \approx 51$
in 2001), this strongly suggests that the observed motions are indeed caused by
the rapid dust evaporation ($v_T \approx v_r > v_{\infty}$) due to backwarming
and higher temperatures in the denser environment formed by the increased mass
loss (see Sect.~\ref{DensTem}).

Furthermore, our new calculations demonstrate that it may not always be
possible to accurately deproject the observed velocities without a
self-consistent radiative transfer modeling. A reason for this complication is
that time-dependent optical depth effects arising from the time-dependent
density distribution (Sect.~\ref{DensTem}) play a major role in the apparent
separations of the observed peaks. This is illustrated by the fact that our
models, while in agreement with the observed distances of component A from the
star (Figs.~\ref{Kimages}, \ref{Knorm}), imply significantly higher radial
velocities of the density peak, $v_{r13} \approx 19$ km\,s$^{-1}$, $v_{r38}
\approx 27$ km\,s$^{-1}$, and $v_{r18} \approx 25$ km\,s$^{-1}$ than simple
geometrical deprojection would indicate (12 km\,s$^{-1}$, 24 km\,s$^{-1}$, and
20 km\,s$^{-1}$, respectively). Possible uncertainties in the distance to
{\irc} cannot decrease the velocities below $v_{\infty}$ even for the lower
limit of 100 pc \citep{Becklin_etal1969}.

\subsubsection{Near-IR color maps}
\label{ColorMaps}

Figure~\ref{Colors} compares the model and observations in terms of the
high-resolution {\JmH}, {\JmK}, and {\HmK} color images for the most recent
epoch of 2001. The speckle color images were computed for only those areas of
the $J$, $H$, and $K$ images, where intensities are greater than 2{\%} of the
peak in the respective bands, whereas the limit for model color images was
1{\%}, to display slightly larger areas. The obvious difference between the
general appearance of the model and speckle images is mainly due to the
inhomogeneities of the intensity distributions in the latter and a large
contribution of components C and D that are very prominent at the last epochs,
but not included in our model. As a consequence, the observed images have
higher intensity levels over larger areas. One can meaningfully compare only
the regions around the bright southern cavity at the latest epochs, in contrast
to the {\HmK} image of 1997 \citepalias[][]{Men'shchikov_etal2001}.

The color images corroborate the conclusion of
\citetalias{Men'shchikov_etal2001} that the bright components A, C, and D are
respectively the cavity and smaller-scale inhomogeneities in the dense shell,
{\em not} dense clumps of dust as it might appear from the images alone
(Fig.~\ref{Kimages}). All the bright components are situated in the blue areas
in both the model and the observed color images. The cavity A coincides with
the bluest spot, which is a natural consequence of lower optical depths along
those directions from the star. The bluest, optically thinnest spot is located
precisely inside the conical cavity, at the surface of the dense dusty shell
close to the dust formation radius. Naturally, the star is in the red area of
the color images, as it was also in the previous epochs \citepalias[Fig.~16
in][]{Men'shchikov_etal2001}. The distribution of red and blue color in the
observed images would be impossible to reproduce in any reasonable and accurate
model, if the star were at the position of the bright southern lobe.

\subsubsection{Shapes of the cavities}
\label{CavityShapes}

The star is clearly visible in the $K$-band model images at the center of the
faint circular halo of dust emission from the hottest and densest inner region
of the dusty envelope (in 1995 and 1997, Fig.~\ref{Kimages}). The brightest
lobe of the southern cavity appeares near the edge of the circular halo, just
at the dust formation zone indicated by two dashed circular segments. The
intensity level of the halo is somewhat higher in the models than in the
observed images (cf. Fig.~\ref{Knorm}), causing its prominent appearance. The
halo is present also in the 1997 $K$ image, where one can see the incompletely
circular halo east of the star. As discussed in
\citetalias{Men'shchikov_etal2001}, the halo is distorted in this image, most
likely due to density inhomogeneities and patchy circumstellar extinction north
and north-west of the star. The 2001 $K$ image is qualitatively similar to the
older image obtained in 1997 but strongly distorted by the merging of the
bright lobe A with component C. Direct stellar light has been absorbed in the
denser dust shell having much larger optical depth ($\tau_{V8} \approx 51$ vs.
$\tau_{V3} \approx 36$).

The shape of the cavity has changed between 1995 and 2001 primarily because of
the brightening of component C and its merging with A. The latter is somewhat
more elongated in the model images in the direction orthogonal to the symmetry
axis (A-B) of the object, which may indicate that the density distribution or
dust properties in the cavities in the model differ from the actual parameters
in {\irc}. Despite somewhat different resolutions of the images, the size
(opening angle) of the outfow cavity seems to be larger in 1995 than at later
epochs, although for 2001 the comparison is complicated by the merger of A and
C. Our modeling suggests that the cross-section of the cavity has indeed
decreased with time, when compared to the cross-section of a cone with fixed
opening angle. Constrained by the observed $K$-band speckle images, the present
models have the opening angle $\omega$ of the conical cavity decreasing from
46{\degr} in 1995 to 36{\degr} in 1997 and to 26{\degr} in 2001. We have been
unable to find models for the first and the last epochs with the same,
unchanged $\omega = 36${\degr}, that would reproduce intensity ratios of the
peaks A and B in the observed images.

This decrease of $\omega$ may be interpreted as a kind of ``collimation'' of
the bipolar cavities, although one must bear in mind that the ``cavities'' in
reality are not necessarily conical, as the model assumes for simplicity.
Moreover, the details of the intensity distribution in the model images are
not completely similar to what is shown by observations (Fig.~\ref{Knorm}). The
model predicts component A to have wider profiles than those of our speckle
images, suggesting that the density distribution inside the cavities, in
contrast to the model assumptions, may depend on the polar angle or be even
more complex.

\begin{figure}
\begin{center}
\resizebox{\hsize}{!}{\includegraphics{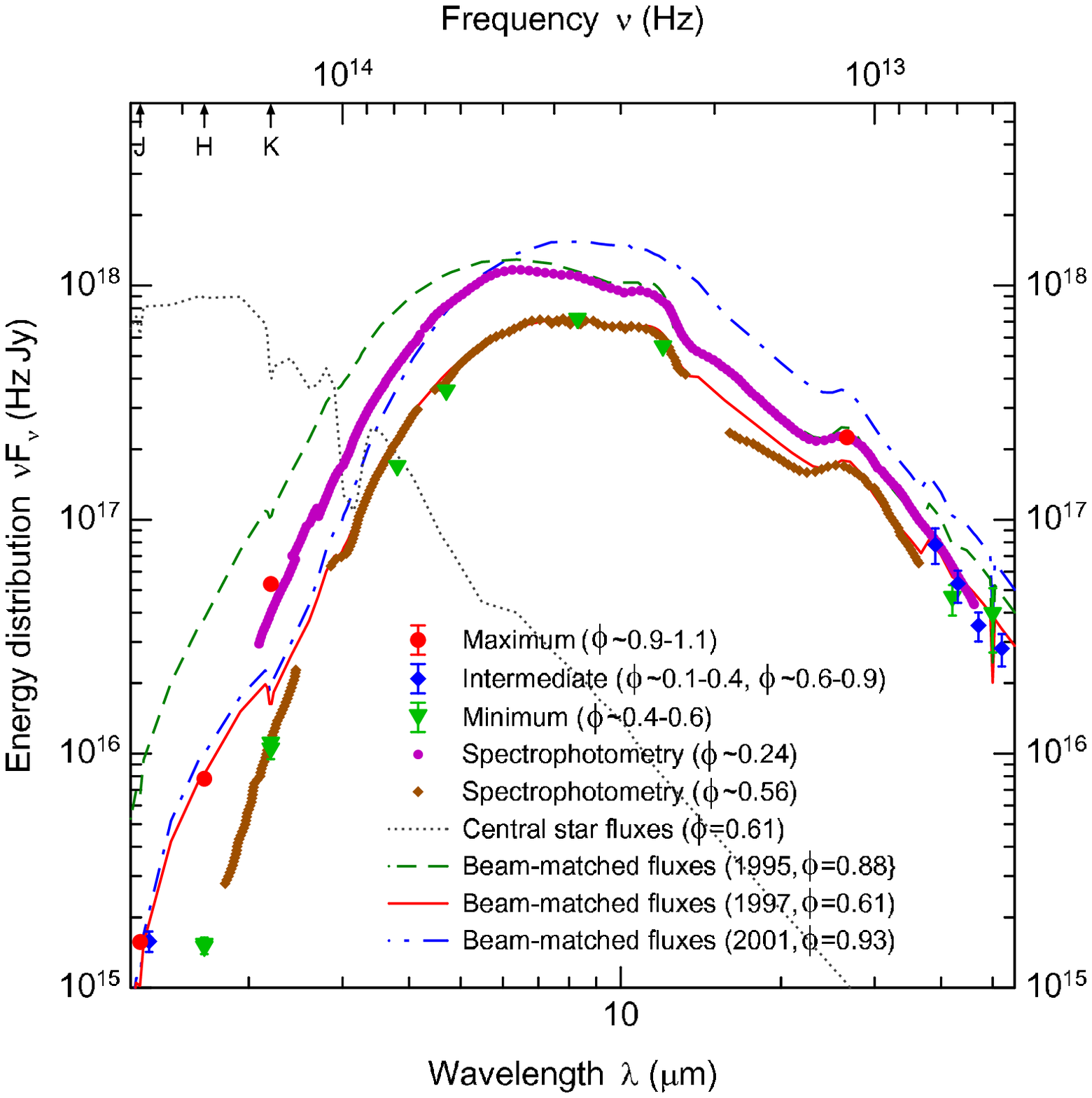}}
\caption
{
Top portions of the model SEDs corresponding to the epochs of 1995, 1997, and
2001 compared to the observed fluxes of {\irc}
\citepalias{Men'shchikov_etal2001}. The beam-matched fluxes of the bipolar
envelope at all the epochs are plotted, as well as the stellar continuum (for
1997). The observed broad-band fluxes are shown by different symbols for
various phases indicated in the legend. If available, error bars for the fluxes
are drawn, whenever they are larger than the symbols. Thick curves (dots) show
all available spectrophotometry data. The model assumes that we observe {\irc}
at the viewing angle ${\theta}_{\rm v}$ = 40{\degr} (below the midplane). Three
arrows at the top abscissa indicate the central frequencies of the $J$, $H$,
and $K$ photometric bands.
}
\label{Continuum}
\end{center}
\end{figure}


\subsection{Spectral energy distribution}
\label{SED}

Figure~\ref{Continuum} compares the model SED with available continuum fluxes
of {\irc} \citepalias[described in][]{Men'shchikov_etal2001}. Since the central
energy source is a pulsating star, its luminosity, effective temperature, and
radius are also changing with a period of 649 days (Sect.~\ref{Parameters}).
Our images documented the envelope's dynamic evolution over 5.4 years, when the
material had only time to move by no more than 16 AU with the wind velocity of 15
km\,s$^{-1}$. Thus, the observed changes have affected only the innermost hot
regions of the envelope ($r \la 100$ AU, $T_{\rm d} \ga 300$ K) and not
influenced the SED at $\lambda \gg 10$ {\mic}. For this reason, we display in
Fig.~\ref{Continuum} only the spectral region 1--50 {\mic} around the
continuum peak. The entire SED and the prominent dust features at 11.3 {\mic}
and 27 {\mic} are discussed in detail in \citetalias{Men'shchikov_etal2001}.

The resulting SED displays large phase variations and significant non-periodic
evolution of the star and the inner dense envelope of {\irc}. Our model shows
that, despite the rapid expansion of the dust formation zone, the visual optical
depth of the dense shell has increased from ${\tau}_{V1} \approx 30$ to
${\tau}_{V3} \approx 37$ to ${\tau}_{V8} \approx 51$ in 1995--2001, which
implies a significant dust density enhancement in the innermost dust formation
region (see Sect.~\ref{DensTem}). The model predicts that the continuum fluxes
at $\lambda \la 5$ {\mic} are now lower than in 1995 by a factor of $\sim 1.5$
and that at $\lambda \ga 5$ {\mic} they are now higher by approximately the
same factor. The model continuum peak shifted toward longer wavelengths from
6.4 {\mic} to 8.6 {\mic} as the increasingly dense and opaque dust formation
zone expanded radially by a factor of 2 (Fig.~\ref{Kimages},
Sect.~\ref{DensTem}). The dust emission features at 11.3 {\mic} and 27 {\mic}
have now become noticeably shallower than several years ago. These model
predictions can, in principle, be tested by spectrophotometric measurements.

\begin{figure*}
\begin{center}
\resizebox{\hsize}{!}
   {
    \hspace{0mm}
    \includegraphics{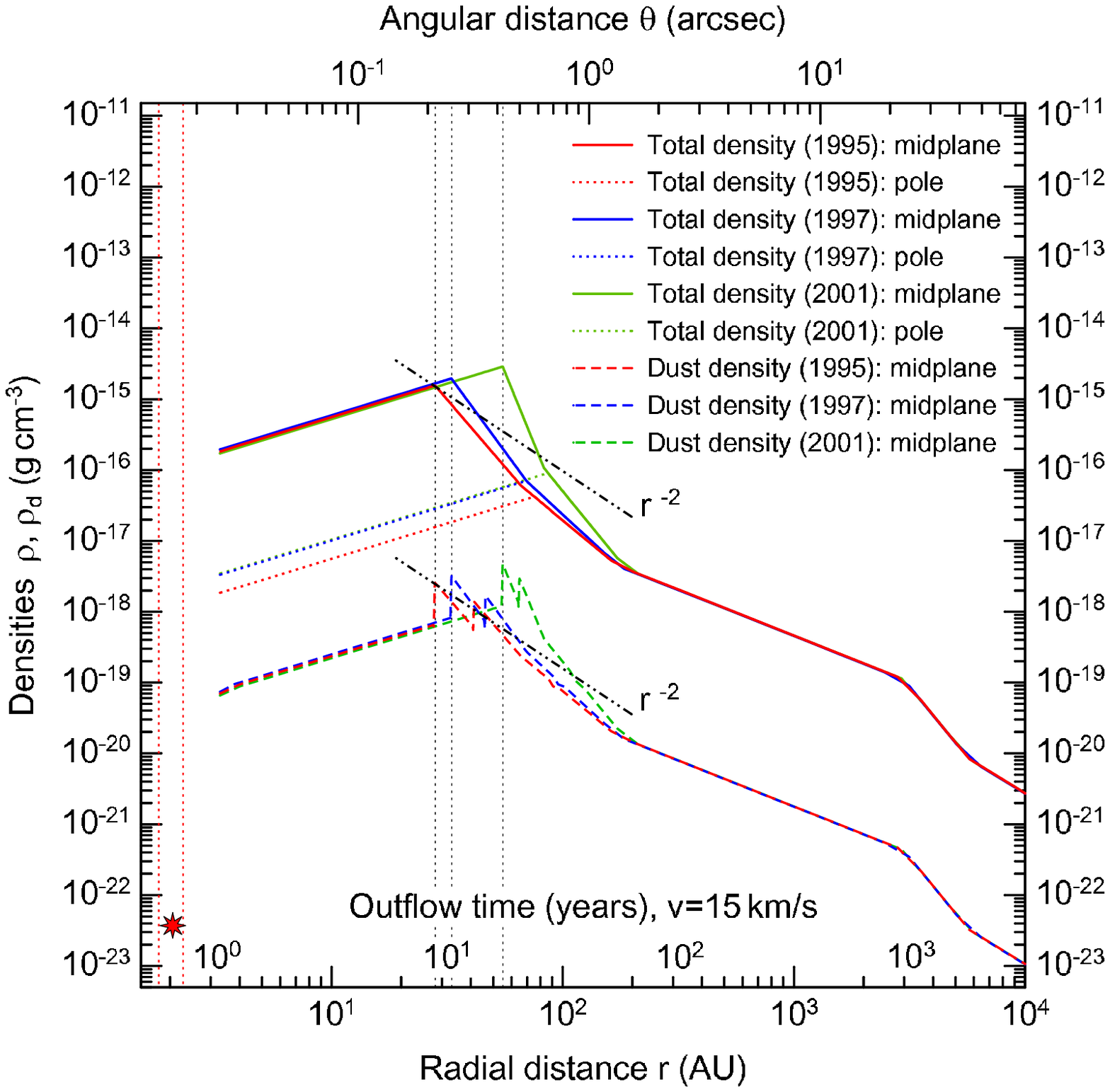}
    \hspace{5mm}
    \includegraphics{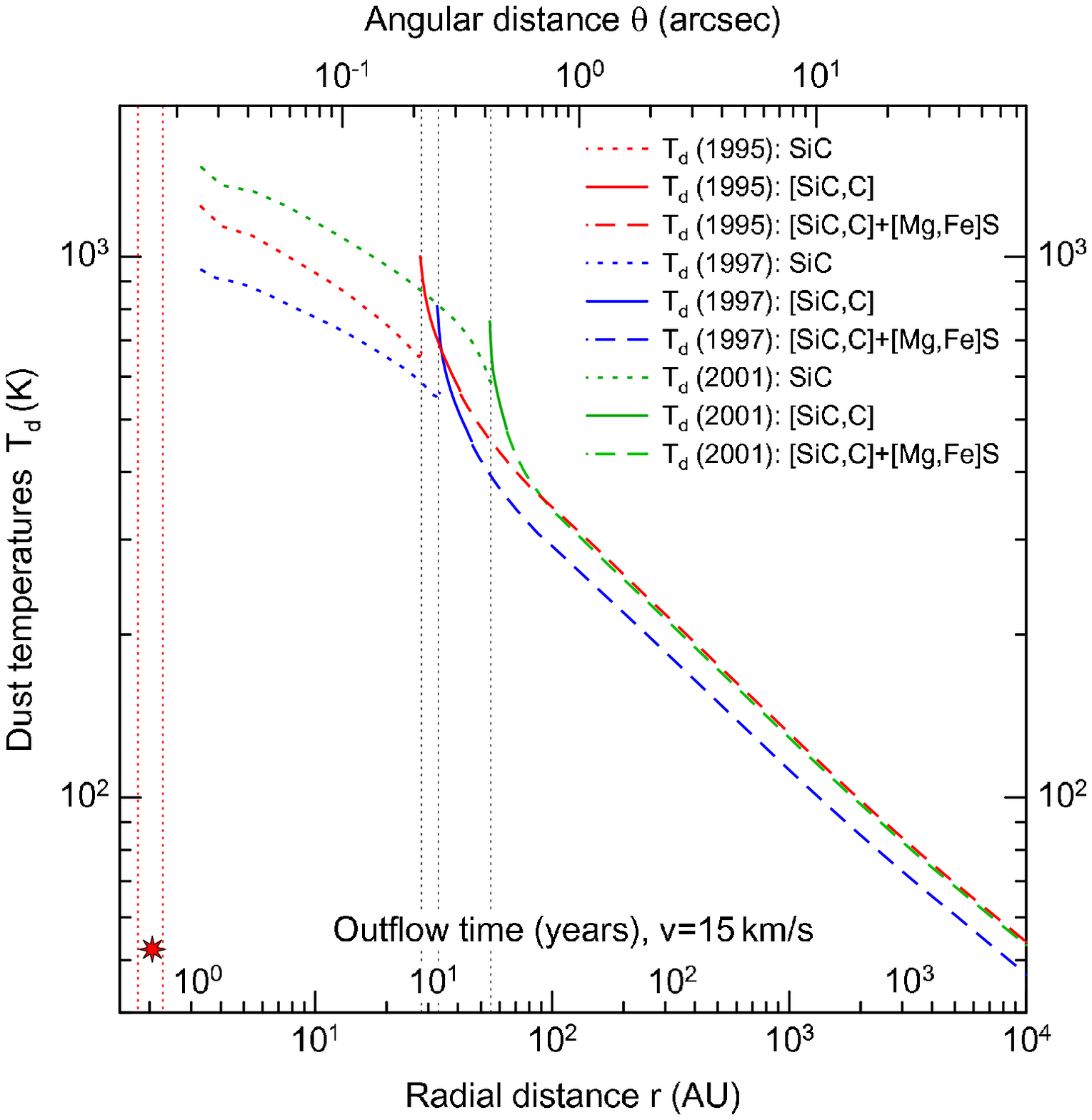}
   }
\caption
{
Model densities ({\em left panel}) and dust temperatures ({\em right panel}) in
the inner envelope of {\irc}. Dust densities and temperatures are displayed for
the smallest grains only ($a = 0.01$ {\mic}) of all dust components. Both
total (gas+dust) density and dust density distributions in the midplane and in
bipolar cavities along the symmetry axis are shown here. The total density
profiles were plotted under the assumption that {\dustgas} of different dust
components do not change with time. Three vertical dotted lines in the middle
of the plot mark the amorphous carbon condensation zones in 1995, 1997, and
2001. Jumps in densities and temperatures in the area are due to sequential
formation of new dust components \citepalias{Men'shchikov_etal2001}.
Dash-double-dotted lines show a $\rho \propto r^{-2}$ density profile that
would form in the case of constant $\dot{M}$ and $v$. The additional labeling
above the lower abscissa shows expansion times assuming a constant outflow
velocity $v = 15$ km\,s$^{-1}$. The two vertical lines drawn on the left side
of the diagrams (labeled with an asterisk) indicate stellar radii at the
luminosity phases $\phi = 0.0$ and $\phi = 0.5$.
}
\label{DenTem}
\end{center}
\end{figure*}


\subsection{Densities and temperatures}
\label{DensTem}

Figure~\ref{DenTem} displays density and temperature distributions in our
models of {\irc} for 1995, 1997, and 2001. The plots show only the inner
regions ($r < 10^4$ AU) of its huge envelope having an outer radius of $6
\times 10^5$ AU; for a discussion of the density and temperature structure of
the entire envelope, we refer to \citetalias{Men'shchikov_etal2001}. The dust
temperatures in Fig.~\ref{DenTem} are plotted for only the smallest dust grains
of the components considered in our dust model. The latter consists of SiC,
aggregates composed of silicon carbide and amorphous carbon [SiC,C], and the
same aggregates covered with the mantles of magnesium-iron sulfides
[SiC,C]+[Mg,Fe]S \citepalias{Men'shchikov_etal2001}.

The density profile of our previous model was adjusted so as to produce the
$K$-band images consistent with the observed speckle images of {\irc}
(Figs.~\ref{Kimages}, \ref{Knorm}). The resulting density distribution shown in
Fig.~\ref{DenTem} changes with time due to the increasing mass-loss rate and
forming dust in the dense shell expanding away from the central star. Between
the epochs, the main dust formation zone displaced from $r_1 \approx$ 28 AU to
$r_3 \approx 33$ AU to $r_8 \approx 55$ AU. Its linear expansion velocity in
our model, $v_{r38} \approx 27$ km\,s$^{-1}$, is almost twice larger than the
general outflow speed of the envelope (Sect.~\ref{Motion}).

The total density distributions shown in Fig.~\ref{DenTem} assume that the
dust-to-gas mass ratios {\dustgas} of different dust components in {\irc} do
not change in time and that they are equal to the values adopted in our
previous time-independent model \citepalias[see Table~2
in][]{Men'shchikov_etal2001}. An inspection of Fig.~\ref{DenTem} shows that the
densities in the dust formation zones of different dust materials became higher
with time, implying that the mass-loss rate $\dot{M}$ increased, too. For an
illustration, we plotted $\rho \propto r^{-2}$ density profiles through
the dust and total density peaks of 1995. These lines would trace the
displacement of the peaks with time, if $\dot{M}$, $v$, and {\dustgas} remained
constant. Contrasting with the expected density distribution of such an
outflow, the model densities rise significantly above the pure expansion line.
This can be interpreted in terms of the increasing mass-loss rate or of the
increasing degree of dust condensation or both.

In the standard picture of mass-losing AGB stars, the radiation pressure
acceleration of dust and gas takes place in the dust formation zone. In
reality, different dust components form within a considerable range of radial
distances from the star \citepalias[cf.][]{Men'shchikov_etal2001}, although the
general picture of the acceleration of the newly-formed dust grains due to the
radiation pressure remains valid. The velocity $v_{r38} \approx 27$ km\,s$^{-1}$
of the wind material close to the dust formation zone, which is higher than the
generally accepted value of $v_{\infty} \approx 15$ km\,s$^{-1}$, does not fit
readily into this context, as it would imply significant deceleration of the
wind material. Although such a deceleration may well be possible in a complex,
time-dependent outflow hydrodynamics, in Sect.~\ref{Motion} we have interpreted
this fast expansion as not the real outflow motion, but rather a displacement
of the dust formation radius due to the backwarming effect in the optically
thicker environment.

In fact, the radial dust temperature distributions of our model
(Fig.~\ref{DenTem}) demonstrate a very significant heating of the grains within
the dense shell. The model density profile just outside the carbon nucleation
zone steepens with time from $\rho_{\rm d1} \propto r^{-3.8}$ to $\rho_{\rm d3}
\propto r^{-4.5}$ to $\rho_{\rm d8} \propto r^{-8}$ between the three
epochs, causing higher values and much steeper distributions of dust
temperatures in that region. When they jump over the condensation temperature
of the dust material, evaporation of the dust grains is a natural and
inevitable consequence. Dust sublimation can quickly shift the dust formation
zone outward, thus bringing the increased densities in balance with the
condensation temperatures of carbon dust, $T_{\rm C} \sim 10^3$ K, however, at
a larger radius. Such displacement of the dust density peak due to the
self-regulating processes of dust formation, heating, and destruction can
easily be mistakenly interpreted as a radial expansion of the wind material.


\subsection{Mass-loss rate and dust formation}
\label{MassLoss}

\begin{figure}
\begin{center}
\resizebox{0.99\hsize}{!}{\includegraphics{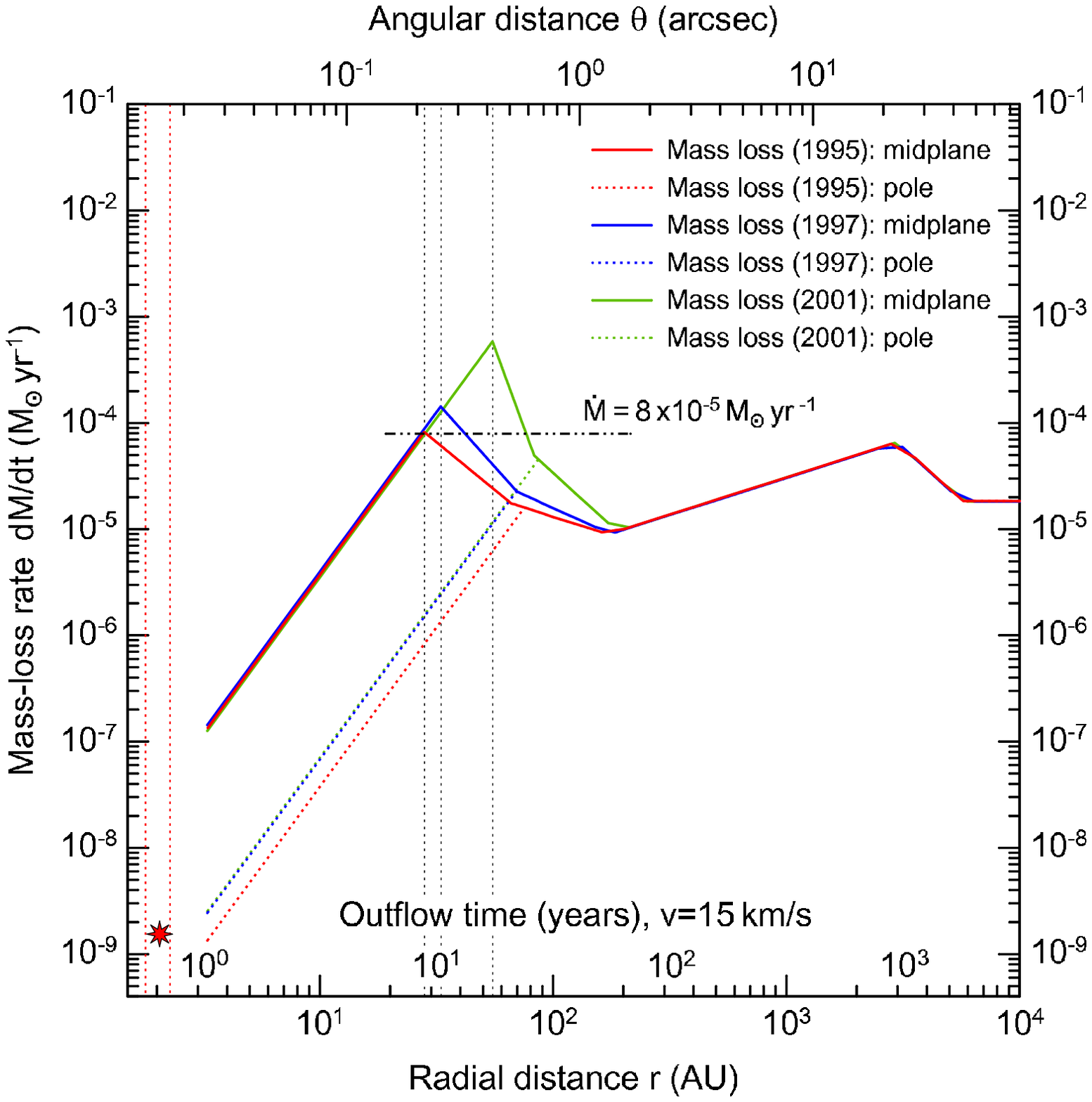}}
\caption
{
Mass-loss history of the inner envelope of {\irc} for October 1995, January
1997, and March 2001, derived from the total density distributions displayed in
Fig.~\ref{DenTem}. The total mass-loss rates in the midplane and along the
model symmetry axis (in polar cavities) are plotted, assuming that the outflow
velocity $v = 15$ km\,s$^{-1}$ is constant throughout the envelope and
that {\dustgas} of all dust components does not change with time. The
dash-double-dotted line corresponds to a constant $\dot{M}$ spherical outflow
resulting from the $\rho \propto r^{-2}$ density profile (Fig.~\ref{DenTem}).
}
\label{MLoss}
\end{center}
\end{figure}

The time-dependent mass-loss rate $\dot{M}$ of {\irc} can readily be derived
from the density distribution of our models, assuming that the outflow velocity
$v$ = 15 km\,s$^{-1}$ is constant across the envelope and that {\dustgas}
in the envelope does not change with time (Fig.~\ref{MLoss}). Note that the
radiative transfer models like this predict only {\em dust} densities and
convert them into gas densities by adopting a specific value (usually arbitrary)
of the dust-to-gas mass ratio. However, {\dustgas} is not a free parameter
in our model of {\irc} \citepalias[see][]{Men'shchikov_etal2001}.

If we assume that {\dustgas} remained constant during the envelope's expansion,
then the rising densities displayed in Fig.~\ref{DenTem} would imply
monotonically increasing mass-loss rates illustrated in Fig.~\ref{MLoss}. In
this case, our model predicts that the mass-loss rate has increased by a factor of
7 during the 5.4 years of our imaging, from $\dot{M}_{1} \approx 8 \times 10^{-5}$
{\Msun}\,yr$^{-1}$ in 1995 to $\dot{M}_{3} \approx 1.4 \times 10^{-4}$
{\Msun}\,yr$^{-1}$ in 1997 to $\dot{M}_{8} \approx 5.8 \times 10^{-4}$
{\Msun}\,yr$^{-1}$ in 2001. Such changes would amount to a rate of $\sim
10^{-4}$ {\Msun}\,yr$^{-2}$. Assuming that the latter is constant and that it
takes 20 years for the material leaving the star to reach the zone of carbon
dust formation (at $r_8 \approx$ 55 AU, Fig.~\ref{DenTem}), we would conclude
that the mass-loss rate from the stellar surface may have been recently as high
as $2.6 \times 10^{-3}$ {\Msun}\,yr$^{-1}$.

If we assume instead that $\dot{M}$ remained constant during the envelope's
expansion, then the total density profiles must differ from those plotted in
Fig.~\ref{DenTem}. The total density peak of 1995 should evolve then along the
$\rho \propto r^{-2}$ track. This would imply that {\dustgas} has increased by
the same factor of 7, from $\rho_{\rm d1}/\rho \approx 9.4 \times 10^{-4}$ in
1995 to $\rho_{\rm d3}/\rho \approx 1.65 \times 10^{-3}$ in 1997 to $\rho_{\rm
d8}/\rho \approx 6.6 \times 10^{-3}$ in 2001, i.e. that the new dust formation
continues as the dense gas flows outward. This seems to be natural, since the
dust formation process takes considerable time \citepalias[see,
e.g.,][]{Weigelt_etal2002}. In this case, the mass-loss rate would remain at a
lower value of $8 \times 10^{-5}$ {\Msun}\,yr$^{-1}$ but {\dustgas} would be
rather high at present, approaching 0.7{\%}.

It seems very likely, however, that in {\irc} both $\dot{M}$ and {\dustgas}
are increasing with time. If this indeed is the case, their values
can be not as high as in the above two scenarios. For example,
if their contributions were equal, then we would derive $\dot{M}_8 \approx 3.3
\times 10^{-4}$ {\Msun}\,yr$^{-1}$ and $\rho_{\rm d8}/\rho \approx 0.004$.

\section{Conclusions}
\label{Conclusions}

We presented results of two-dimensional radiative transfer modeling of {\irc}
at selected moments of its recent evolution, which correspond to three epochs
of our series of high-resolution speckle images of the object recorded over the
last 5.4 years \citepalias{Osterbart_etal2000,Weigelt_etal2002}. The imaging
revealed dynamic evolution of the subarcsecond dusty environment of {\irc} on a
very short time scale, of the order of one year. The goal of the present study
was to test predictions of our recent time-independent modeling
\citepalias{Men'shchikov_etal2001} which derived the structure and physical
properties of {\irc} at a single epoch (January 1997). Now that the entire
sequence of 8 high-resolution near-IR images has documented complex changes in
the inner bipolar dusty shell of the carbon star, we took these temporal
constraints into account to see, whether the static model is consistent with
the fast {\em evolution} seen in our images. The new modeling allowed us to
make a quantitative physical interpretation of what goes on in {\irc}.

Our previous static model of {\irc} \citepalias{Men'shchikov_etal2001} is
consistent with the new constraints, provided that small modifications are made
to its density profile in the dust formation zone and to the opening angle of
the bipolar cavity. The new modeling has shown that the cavity has been
shrinking from $\omega_1 \approx 36${\degr} to $\omega_8 \approx 26${\degr} and
the density distribution across it has been changing during the 5.4 years of
our imaging. We are witnessing a dynamic episode of rising mass loss from the
central star, from $\dot{M} \approx 10^{-5}$ {\Msun}\,yr$^{-1}$ to the present
value of $\dot{M} \approx 3 \times 10^{-4}$ {\Msun}\,yr$^{-1}$, which probably
started $\sim$ 50 years ago. If the current rate of the increase of
$\dot{M}$ is constant, the mass-loss rate from the stellar surface may be now
as high as $2.6 \times 10^{-3}$ {\Msun}\,yr$^{-1}$.

A compact dense shell with bipolar cavities has formed around the star as a
result of the rapid increase of the mass loss by {\irc}, causing the observed
rapid evolution documented in our images. The higher mass loss produces
favorable conditions for dust formation in the increasingly dense inner
envelope expanding outward with the outflow velocity $v \approx 15$
km\,s$^{-1}$. Larger amounts of dust increased the optical depths, obscuring the
central star, whereas the optically thinner cavity remained relatively
unaffected. Due to backwarming, temperatures in the dust formation zone became
higher, shifting the latter to larger radii at the velocity $v_T \approx 27$
km\,s$^{-1}$ and mimicking the real outflow motion of the circumstellar shell
material with that speed. One can predict that the star will remain obscured
until $\dot{M}$ starts to drop back to lower values in the dust formation zone.
Within a few years from that moment, we could be witnessing the star
reappearing, whereas the cavities becoming relatively fainter.




\bibliographystyle{aa}
\bibliography{aamnem99,h3721}
\end{document}